\documentclass[10pt,journal]{IEEEtran}
\usepackage{amsmath,amssymb,amsfonts}
\usepackage{graphicx}
\usepackage{booktabs}
\usepackage{cite}
\usepackage{algorithm}
\usepackage{algpseudocode}
\usepackage[hidelinks]{hyperref}

\newtheorem{theorem}{Theorem}
\newtheorem{proposition}{Proposition}
\newtheorem{remark}{Remark}
\newtheorem{assumption}{Assumption}
\newtheorem{definition}{Definition}
\newtheorem{problem}{Problem}
\newtheorem{lemma}{Lemma}
\newtheorem{corollary}{Corollary}

\newcommand{\xb}{\bar{x}}
\newcommand{\ub}{\bar{u}}
\newcommand{\R}{\mathbb{R}}
\newcommand{\E}{\mathbb{E}}

\newif\ifanonymous
\anonymousfalse 

\title{CS-MPC: Contraction-Certified Sampling-Based MPC for Nonlinear Systems Under Uncertainty}
\ifanonymous
\author{Anonymous Authors}
\else
\author{Yashwanth Kumar Nakka%
\thanks{The author is with the Daniel Guggenheim School of Aerospace Engineering, Georgia Institute of Technology, Atlanta, GA 30332 USA (e-mail: \texttt{ynakka3@gatech.edu}).}}
\fi
\usepackage{xcolor}
\newcommand{\rev}[1]{{#1}} 

\begin{document}
\maketitle

\begin{abstract}
We present contraction-certified sampling-based model predictive control (CS-MPC) for tracking and safety of nonlinear robotic systems under parametric uncertainty, disturbances, and safety constraints that are revealed online. CS-MPC uses a contraction-certified tracking feedback inside every sampled rollout, which concentrates the rollouts around the reference, and an execution layer that verifies the applied input using a quantile-tightened discrete-time control barrier function condition. We derive a verification procedure for nonlinear contraction of the feedback-closed dynamics, and we prove a finite-sample stochastic stability bound and a mission-level probabilistic safety guarantee. We validate CS-MPC in simulation on a rover, a quadrotor, and an asteroid landing. When compared with standard model predictive path integral (MPPI) control, CS-MPC attains the same tracking error with 7 to 20 times fewer rollouts for the rover, and a 2 to 4 times lower tracking error for the quadrotor, with no safety violations in 200 Monte Carlo missions.
\end{abstract}


\section{Introduction}
\label{sec:introduction}

\begin{figure}[t]
\centering
\includegraphics[width=0.85\columnwidth]{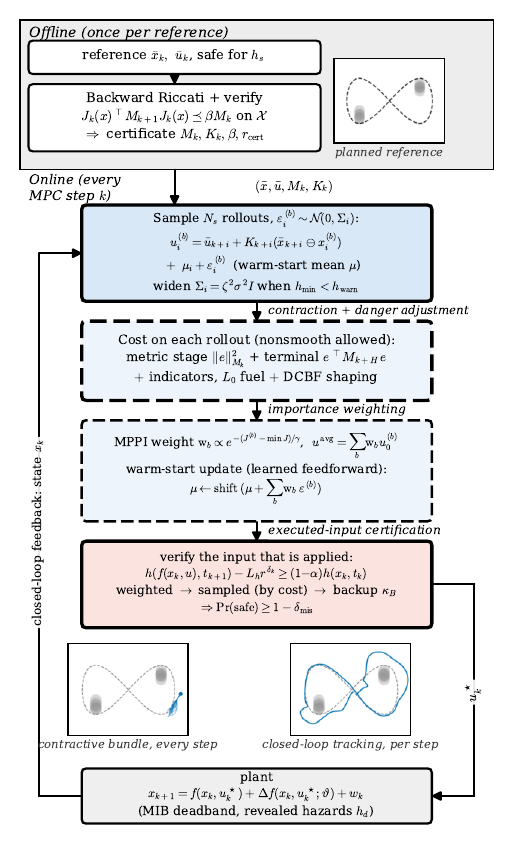}
\caption{Overview of CS-MPC. Offline, a Riccati recursion synthesizes a time-varying metric and gain along the reference, and the contraction of the exact nonlinear closed loop is verified over an explicit domain. Online, the \emph{stability and performance layer} uses the contraction-certified tracking feedback inside every sampled rollout, which concentrates the rollouts around the reference (inset), while the MPPI weights evaluate nonsmooth mission objectives directly, and the \emph{execution layer} verifies the applied input using a quantile-tightened discrete barrier condition, with sampled and backup fallbacks. The system is affected by a mission parameter $\vartheta$ drawn once per run, process noise, actuator quantization, and constraints revealed during the mission.}
\label{fig:overview}
\end{figure}

Planetary rovers~\cite{maimone2007,thakker2023eels}, spacecraft operating near small bodies~\cite{scheeres2012}, and agile aerial robots~\cite{szmuk2017convexification} must track a planned trajectory in environments that are only partially known before the mission. A rover follows a path over terrain with uncertain slip. A spacecraft descends toward a small body with uncertain gravity and spin. An aerial robot flies among moving obstacles that are detected only during flight. In each case, the dynamics are nonlinear, some model parameters are uncertain for the whole mission, the actuators have deadband or quantization, and some safety constraints are revealed only online. Sampling-based model predictive control (MPC)~\cite{kappen2005,williams2016,williams2017} computes the control input from Monte Carlo rollouts of the dynamics, which allows nonsmooth and nonconvex mission costs to be optimized without gradients. However, standard model predictive path integral (MPPI) control~\cite{williams2018information} has three limitations for these problems. First, the rollouts are propagated in open loop, so they spread quickly when the dynamics are unstable, and many samples are needed for a useful control update. Second, the safety constraints enter as rollout penalties, which shape the sampling distribution but do not certify the input that is applied to the system. Third, the effect of a finite number of samples on the closed-loop tracking error is usually evaluated only in simulation.

In this work, we present contraction-certified sampling-based MPC (CS-MPC), which addresses these three limitations using two layers, as shown in Fig.~\ref{fig:overview}. The stability and performance layer computes a tracking feedback gain and a contraction metric offline along the reference trajectory using a backward Riccati recursion~\cite{manchester2017,weimccloybao2021,weimccloybao2022}, and verifies that the nonlinear closed-loop dynamics are contracting on a specified domain. CS-MPC uses this feedback inside every sampled rollout, which keeps the rollouts close to the reference, while the MPPI weights optimize the sampled feedforward input for nonsmooth mission costs. The execution layer verifies the weighted input before it is applied, using a discrete-time control barrier function (DCBF) condition~\cite{agrawal2017,zeng2021} that is tightened by a conditional quantile region for the one-step prediction error. If the weighted input does not satisfy this condition, then the execution layer applies a feasible sampled input or a backup policy. The analysis accounts for the persistent parameter uncertainty, the process noise, the actuator error, and the finite number of samples.

The main contributions of this work are as follows:
\begin{enumerate}
\renewcommand{\labelenumi}{\alph{enumi})}
    \item We derive an offline procedure that synthesizes the tracking feedback and verifies nonlinear contraction of the closed-loop dynamics over a specified domain, which gives nominal and stochastic tracking bounds. We prove that the spread of the feedback-closed rollouts is bounded uniformly in the horizon, which is the mechanism behind the sample efficiency observed in simulation.
    \item We present an execution layer that verifies the applied input using a quantile-tightened DCBF condition, which handles nonconvex, nonsmooth, and time-varying safe sets. We prove a mission-level probabilistic safety guarantee when the prediction-error region is conditionally valid and a feasible backup policy is available.
    \item We derive a concentration bound for the weighted control input and use it to prove a finite-sample stochastic stability bound for the closed-loop system, which separates the effects of the sample budget, the disturbances, and the avoidance maneuvers.
    \item We validate CS-MPC in simulation on a rover under terrain-slip uncertainty, a quadrotor flying among moving obstacles, and an asteroid landing with minimum-impulse thrusters. We also validate the stability and concentration bounds on a stationkeeping experiment with bounded disturbances.
\end{enumerate}

\subsection{Related Work}\label{sec:related}
Sampling-based MPC computes the control input from Monte Carlo rollouts of the dynamics. The path-integral and free-energy formulations were derived in~\cite{kappen2005,theodorou2010}, and the receding-horizon MPPI algorithm along with its information-theoretic interpretation was presented in~\cite{williams2016,williams2017,williams2018information}; the underlying importance-sampling estimators are surveyed in~\cite{deboer2005}. Several methods reduce the rollout dispersion or the model mismatch of MPPI. Tube-MPPI~\cite{williams2018tube} tracks the sampled plan with an ancillary iLQG controller, robust MPPI~\cite{gandhi2021} augments the importance sampler with feedback and analyzes the growth of the free energy, and $\mathcal{L}_1$-adaptive augmentation~\cite{pravitra2020} compensates for matched uncertainty below the sampling rate. Smooth MPPI~\cite{kim2022smooth} penalizes the input derivatives, covariance-steering MPPI~\cite{balci2022,yoonkim2026} constrains the terminal rollout dispersion, and gradient-informed or learned proposal distributions~\cite{mohamed2022,bhardwaj2021,chua2018} reduce the required sample budget. These methods reduce the rollout dispersion or the model mismatch, but the tracking feedback is either applied outside the sampled rollouts or computed from local linearizations without a contraction rate over an explicit domain, and the finite-sample error of the weighted input is not propagated to the closed-loop system. We use the tracking feedback inside every sampled rollout and verify its contraction offline for the nominal nonlinear feedback-closed dynamics over an explicit domain, which keeps the rollouts concentrated around the reference for any horizon. We then propagate the finite-sample error of the weighted input into a stochastic stability bound for the closed-loop system, and we verify the applied input using a DCBF condition, which gives a mission-level probabilistic safety guarantee.

Contraction theory~\cite{lohmiller1998,aminzare2014} characterizes incremental stability through the variational dynamics, and control contraction metrics~\cite{manchester2017} extend this analysis to feedback synthesis. In discrete time, the metric can be constructed through Riccati-type recursions~\cite{weimccloybao2021,weimccloybao2022}, and stochastic incremental stability is surveyed in~\cite{tsukamoto2021}. In predictive control, these certificates are used in tube MPC~\cite{mayne2000,mayne2005,kohler2021}, contraction-based tube planning~\cite{singh2017}, tubes with cross sections that are adapted online~\cite{lopez2019}, and funnel libraries for sequential composition~\cite{majumdar2017}, while more recent work learns the certificates from data~\cite{sun2020,zhao2022}. Certified tracking methods that solve a smooth or convexified optimization online cannot use discontinuous costs such as indicators, lookup tables, and $\ell_0$ costs directly. We use a Riccati-based metric to certify nominal nonlinear contraction on a specified domain, and we use importance sampling to evaluate these costs. As we will discuss, the feedback inside the rollouts also bounds their spread and improves the sample efficiency.

Control barrier functions~\cite{ames2017,ames2019} certify forward invariance of a safe set through a pointwise condition on the control input. The discrete-time conditions are derived in~\cite{agrawal2017}, their use within MPC is discussed in~\cite{zeng2021,yin2023}, nonsmooth compositions using minimum and maximum operations are studied in~\cite{glotfelter2017}, and predictive safety filters are presented in~\cite{wabersich2021}. In many sampling-based MPC implementations, the barrier appears as a rollout penalty or as a filter on a planned sequence. The admissible input set of a nonlinear barrier can be nonconvex, so that an importance-weighted average of individually feasible sampled inputs need not be feasible, and a rollout penalty does not certify the executed input. Furthermore, stochastic tightening is often derived for temporally independent disturbances, whereas persistent mission parameters can induce prediction errors that are biased and temporally correlated. We verify the barrier condition on the input that is executed. The condition is tightened by a conditional quantile region for the one-step prediction error, which accounts for the persistent mission parameters. If the weighted input does not satisfy the condition, then the controller applies a feasible sampled input or a backup policy (Assumption~\ref{ass:backup}). Under conditional coverage and backup feasibility, we prove the mission-level bound of Proposition~\ref{prop:executed-safety}, which holds for nonsmooth and nonconvex safe sets.

Stochastic MPC~\cite{mesbah2016} handles chance constraints through analytic or scenario-based tightening, learning-based MPC~\cite{hewing2020,chua2018} propagates model uncertainty that is estimated from data, and chance-constrained trajectory optimization under stochastic dynamics has been solved using generalized polynomial chaos~\cite{nakka2023chance}. These formulations typically compute an open-loop or affine-feedback plan, and their guarantees do not transfer directly when a receding-horizon sampling controller replaces the planned inputs. Many formulations also model temporally independent disturbances, whereas the mission parameters considered in this work persist throughout a trajectory, as in planetary and small-body missions~\cite{scheeres2012,maimone2007,thakker2023eels,vaquero2024eels}. We state the guarantees for the closed-loop sampling controller. The unobserved mission parameter enters through conditional moments, and the fixed reference can be computed by a standard planning pipeline~\cite{hauser2016,morgan2016swarm}.

\subsubsection*{Organization} We state the problem in Section~\ref{sec:problem} and discuss the preliminaries in Section~\ref{sec:prelim}. In Section~\ref{sec:method}, we present the CS-MPC method, and in Section~\ref{sec:main-results}, we derive the concentration and stochastic stability bounds. In Section~\ref{sec:sims}, we validate CS-MPC in simulation on the three examples. Finally, we conclude the paper in Section~\ref{sec:conclusion}.

\section{Problem Description}
\label{sec:problem}
In this section, we describe the stochastic system, the safe set, and the reference trajectory, and we state the tracking and safety problem. We consider the discrete-time nonlinear stochastic system
\begin{equation}
    x_{k+1}=f(x_k,u_k)+\Delta f(x_k,u_k;\vartheta)
    +d_{a,k}(x_k,u_k)+w_k,
    \label{eq:dyn}
\end{equation}
where $x_k\in\R^n$ is the state, $u_k\in\mathcal{U}\subset\R^m$ is the commanded input, $\mathcal{U}$ is nonempty, compact, and convex, and $f$ denotes the continuously differentiable nominal dynamics. The structured discrepancy $\Delta f(x_k,u_k;\vartheta)$ is parameterized by a random variable $\vartheta$ that is drawn at the beginning of a mission and held fixed along the resulting trajectory. It captures persistent uncertainty in quantities such as mass, inertia, aerodynamic coefficients, gravity parameters, and terrain properties, and we set $\Delta f\equiv0$ when no structured uncertainty is present. The term $d_{a,k}(x_k,u_k)$ captures the actuator implementation error, such as deadband or quantization, and is set to zero when such effects are absent.

The process disturbance $\{w_k\}_{k\geq0}$ is a zero-mean i.i.d.\ sequence satisfying
\begin{equation}
    \E[w_k]=0,
    \qquad
    \E[w_kw_k^\top]=W\succeq0,
    \label{eq:disturbance_moments}
\end{equation}
and is independent of $\vartheta$, the past states and disturbances, and the controller sampling perturbations, as formalized in Definition~\ref{def:filtration}. For concreteness, one may take $w_k\sim\mathcal{N}(0,W)$, although the mean-square analysis requires only the zero-mean, independence, and finite-covariance assumptions in~\eqref{eq:disturbance_moments}.

For a measured state $x_k$ and candidate input $u$, we define the total one-step prediction error as
\begin{equation}
    d_{x,k}(u):=\Delta f(x_k,u;\vartheta)+d_{a,k}(x_k,u)+w_k.
    \label{eq:total_prediction_error}
\end{equation}
With the nominal prediction $\hat{x}_{k+1}(u):=f(x_k,u)$, the realized state is $x_{k+1}=\hat{x}_{k+1}(u)+d_{x,k}(u)$. Note that $\{d_{x,k}\}$ need not be temporally independent even though $\{w_k\}$ is i.i.d., because $\vartheta$ remains fixed throughout a mission. Therefore, the tracking and safety certificates use conditional moments and conditional high-probability regions for the total prediction error.

The state must remain in the time-varying safe set $\mathcal{X}_{\mathrm{safe}}(t):=\{x\in\R^n\mid h_c(x,t)\geq0\}$, where $h_c:\R^n\times\R_{\geq0}\to\R$ is continuous and locally Lipschitz in $x$. We write $h_c(x,t):=\min\{h_s(x),h_d(x,t)\}$, where $h_s$ is known during reference generation and $h_d$ is predictable over the online MPC horizon but need not be available offline. Each component may contain nested minimum and maximum operations~\cite{glotfelter2017,zeng2021}, which allows intersections, unions, nonconvex obstacles, corridors, keep-out and keep-in regions, velocity or actuator-state envelopes, and moving-obstacle clearance constraints.

The objective is to track the nominal reference $\{(\xb_k,\ub_k)\}_{k=0}^{N-1}$ with terminal state $\xb_N$, where $\xb_{k+1}=f(\xb_k,\ub_k)$ for $k=0,\ldots,N-1$ and $h_s(\xb_k)>0$ for $k=0,\ldots,N$. This reference is dynamically feasible and statically safe for the nominal model. However, the realized trajectory can become unsafe due to the process disturbances, the persistent model uncertainty, the actuator implementation error, and the dynamic constraints that are not available when the reference is generated. Therefore, CS-MPC combines contraction-certified tracking with online probabilistic safety enforcement.

We define the tracking errors as
\begin{equation}
    e_k:=x_k\ominus\xb_k,
    \qquad
    \delta u_k:=u_k-\ub_k,
    \label{eq:errors}
\end{equation}
where $\ominus$ denotes the state-space difference operator. It reduces to ordinary subtraction on Euclidean coordinates, while angular coordinates are wrapped to $(-\pi,\pi]$. We assume the relevant trajectories remain in a coordinate chart that does not cross an angular branch cut, so $\xb_k\ominus x_k=-e_k$.

At each sampling instant $t_k=k\Delta t$, the controller simulates $N_s$ closed-loop rollouts over $\{k,k+1,\ldots,k+H\}$, evaluates their costs, applies the selected first input $u_k^\star$ over $[t_k,t_{k+1})$, and repeats from the newly measured state. The main problem is formalized as follows.
\begin{problem}[MPC Under Stochastic Uncertainty]
\label{prob:main}
Given the nominal reference $\{(\xb_k,\ub_k)\}_{k=0}^{N-1}$ and $\xb_N$, the current state $x_k$, a prediction of $\mathcal{X}_{\mathrm{safe}}(t)$ over the current MPC horizon, and the stochastic uncertainty model, construct a causal policy sequence $\{\pi_k\}_{k=0}^{N-1}$, with $u_k=\pi_k(x_k)$ computed from $N_s$ rollouts per sampling instant, subject to
\begin{subequations}
\label{eq:probmain}
\begin{align}
    &x_{k+1}=f(x_k,u_k)+\Delta f(x_k,u_k;\vartheta)
      +d_{a,k}(x_k,u_k)+w_k,
    \nonumber\\[-0.2em]
    &\hspace{10em}k=0,\ldots,N-1,
    \label{eq:probmain_dyn}
    \\
    &u_k=\pi_k(x_k)\in\mathcal U,
    \quad k=0,\ldots,N-1,
    \label{eq:probmain_input}
    \\
    &\Pr\!\left(\bigcap_{k=0}^{N}
      \{x_k\in\mathcal X_{\mathrm{safe}}(t_k)\}\right)
    \nonumber\\[-0.2em]
    &\hspace{10em}\geq1-\delta_{\mathrm{mis}},
    \label{eq:probmain_safety}
\end{align}
\end{subequations}
where $\delta_{\mathrm{mis}}\in(0,1)$ is the mission-level risk budget and the probability is taken over $\vartheta$, the process disturbances, any initial-state uncertainty, and the randomness introduced by the sampling controller. The tracking performance of the policy is measured by the expected quadratic cost
\begin{equation}
    \E\!\left[\sum_{k=0}^{N-1}
        \bigl(\|e_k\|_Q^2+\|\delta u_k\|_R^2\bigr)\right],
    \qquad Q\succ0,\ R\succ0,
    \label{eq:probmain_cost}
\end{equation}
with $\|e\|_Q^2:=e^\top Qe$ and $\|\delta u\|_R^2:=\delta u^\top R\delta u$. The mean-square analysis of Section~\ref{sec:main-results} bounds the growth of this cost, and the implemented rollout objective~\eqref{eq:cost} also contains mission, exploration, and barrier-shaping terms.

The policy must satisfy the following certification requirements:
\begin{enumerate}
    \item[(i)] The nominal feedback-closed dynamics satisfy a differential contraction condition~\cite{lohmiller1998,manchester2017}, yielding incremental exponential stability on the certified state-space domain. The implemented finite-sample controller also admits a stochastic stability bound that accounts for the population feedforward maneuver, finite-sample approximation error, process noise, persistent model uncertainty, and actuator implementation error.
    \item[(ii)] The applied input satisfies a discrete-time control barrier function condition~\cite{agrawal2017} tightened using a conditional quantile region for~\eqref{eq:total_prediction_error}. The per-step risks satisfy $\sum_{k=0}^{N-1}\delta_k\leq\delta_{\mathrm{mis}}$. This allocation enforces~\eqref{eq:probmain_safety} when the field used in this condition is constructed directly from the physical safety field; for the lookahead implementation, the guarantee applies to the surrogate set described in Remark~\ref{rem:lookahead}. If the weighted input does not satisfy this condition, the controller applies a sampled or backup input that satisfies the same execution-time condition.
    \item[(iii)] The deviation of the finite-sample weighted input from its conditional population counterpart is bounded explicitly as a function of $N_s$.
\end{enumerate}
\end{problem}
\section{Preliminaries}
\label{sec:prelim}
In this section, we introduce the information structure, the reference and regularity assumptions, the contraction tools, the stochastic DCBF conditions, and the MPPI update that are used by CS-MPC.
\subsection{Information Structure}
\begin{figure}[!t]
\centering
\includegraphics[width=\columnwidth]{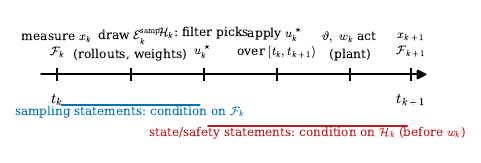}
\caption{Within-step information structure of Definition~\ref{def:filtration}: sampling statements condition on $\mathcal{F}_k$; one-step state and safety statements condition on $\mathcal{H}_k$, after the input is fixed and before $w_k$ is realized.}
\label{fig:timeline}
\end{figure}

\begin{definition}[Controller information and filtrations]
\label{def:filtration}
At MPC step $k$, let $\mathcal{E}_k^{\mathrm{samp}}:=\{\varepsilon_{i\mid k}^{(b)}\mid i=0,\ldots,H-1,\ b=1,\ldots,N_s\}$ denote the perturbations drawn at that step. Let $\mathcal{Z}_k$ be the $\sigma$-algebra generated by the exogenous information available before sampling, including the reference window, predicted constraint fields, and current stochastic-uncertainty model, with $\mathcal{Z}_k\subseteq\mathcal{Z}_{k+1}$. Define the pre-sampling information by
\begin{equation}
\begin{aligned}
    \mathcal{F}_k:=\sigma(&x_0,\ldots,x_k,\ u_0,\ldots,u_{k-1},\\
    &\mathcal{E}_0^{\mathrm{samp}},\ldots,
    \mathcal{E}_{k-1}^{\mathrm{samp}})\vee\mathcal{Z}_k.
\end{aligned}
\end{equation}
The state $x_k$, certificate $(M_k,K_k)$, warm-start mean, and uncertainty model at step $k$ are $\mathcal{F}_k$-measurable. The post-sampling information is $\mathcal{H}_k:=\mathcal{F}_k\vee\sigma(\mathcal{E}_k^{\mathrm{samp}})$, so $\mathcal{F}_k\subseteq\mathcal{H}_k\subseteq\mathcal{F}_{k+1}$. It additionally fixes the rollouts, costs, weights, selected branch of Algorithm~\ref{alg:csmc}, and applied input.

The mission parameter $\vartheta$ is unobserved, and its conditional law given $\mathcal{F}_k$ represents the remaining structured uncertainty. The fresh disturbance $w_k$ is independent of $\mathcal{H}_k\vee\sigma(\vartheta)$ and satisfies $\E[w_k\mid\mathcal{H}_k]=0$ and $\E[w_kw_k^\top\mid\mathcal{H}_k]=W$. Conditioned on $\mathcal{F}_k$, the perturbations in $\mathcal{E}_k^{\mathrm{samp}}$ are independent across $i$ and $b$ and independent of $\vartheta$, $w_k$, and future process disturbances.
\end{definition}
We use two conditioning conventions in the following. The sampling statements are conditioned on $\mathcal{F}_k$, which is before the perturbations are drawn. The one-step state and safety statements are conditioned on $\mathcal{H}_k$, which is after the sampled input is fixed and before $w_k$ is realized, so that the process-noise cross terms vanish. The total prediction error can still have a nonzero conditional mean, because it also contains the persistent model discrepancy and the actuator implementation error. The timeline within one step is shown in Fig.~\ref{fig:timeline}.
\subsection{Reference Trajectory and Regularity}
\begin{assumption}[Reference trajectory]
\label{ass:ref}
A nominal reference $\{(\xb_k,\ub_k)\}_{k=0}^{N-1}$ and terminal state $\xb_N$ are given such that $\xb_{k+1}=f(\xb_k,\ub_k)$ and $\ub_k\in\mathcal{U}$ for $k=0,\ldots,N-1$, while $h_s(\xb_k)\geq m_s>0$ for $k=0,\ldots,N$.
\end{assumption}
In the simulation studies of Section~\ref{sec:sims}, the reference is generated offline by a stabilized nominal rollout, by differential flatness, and by a quintic descent with nominal inverse dynamics. In cluttered environments, the reference can instead be computed using asymptotically optimal kinodynamic planning~\cite{hauser2016} followed by sequential convex programming~\cite{morgan2016swarm}.
\begin{assumption}[Regularity and Stabilizability]
\label{ass:regularity}
Let $\mathcal{X}$ denote the certified state-space domain, including the appropriate angular topology.
\begin{enumerate}
    \item[(a)] The nominal dynamics $f$ are $C^2$ on an open neighborhood of $\mathcal{X}\times\mathcal{U}_{\mathrm{ext}}$, where $\mathcal{U}_{\mathrm{ext}}$ contains the nominal feedback values used in the contraction analysis.
    \item[(b)] The reference linearization pair $A_k:=\partial_x f(\xb_k,\ub_k)$ and $B_k:=\partial_u f(\xb_k,\ub_k)$ is uniformly stabilizable.
    \item[(c)] Let $A(x,u):=\partial_x f(x,u)$, $B(x,u):=\partial_u f(x,u)$, and $u_k^{\mathrm{fb}}(x):=\ub_k+K_k(\xb_k\ominus x)$. Define
    \begin{subequations}
    \begin{align}
        F_k(x)
        &:={}
        f\!\left(x,u_k^{\mathrm{fb}}(x)\right),
        \label{eq:closed_loop_map}
        \\
        J_k(x)
        &:={}
        \partial_xF_k(x)
        \nonumber\\
        &\hphantom{:={}}=A\!\left(x,u_k^{\mathrm{fb}}(x)\right)
        -B\!\left(x,u_k^{\mathrm{fb}}(x)\right)K_k.
        \label{eq:closed_loop_jacobian}
    \end{align}
    \end{subequations}
    At the reference, $\Phi_k:=J_k(\xb_k)=A_k-B_kK_k$.
    \item[(d)] A verified bound $\Delta_{\Phi,k}\geq0$ is available such that
    \begin{equation}
        \sup_{x\in\mathcal{X}}\|J_k(x)-\Phi_k\|_2
        \leq\Delta_{\Phi,k},
        \qquad k=0,\ldots,N-1.
        \label{eq:jacobian_deviation_bound}
    \end{equation}
\end{enumerate}
The bound may be obtained analytically or through verified interval arithmetic, automatic differentiation, or branch-and-bound. We use random gridding for numerical validation of the bound.
\end{assumption}
\begin{assumption}[Conditional stochastic prediction error]
\label{ass:stochastic-prediction-error}
For every $\mathcal{H}_k$-measurable $u\in\mathcal{U}$, the total prediction error~\eqref{eq:total_prediction_error} has finite conditional second moment. With $\widetilde d_{x,k}(u):=d_{x,k}(u)-\bar d_{x,k}(u)$, define
\begin{equation}
\begin{aligned}
    \bar d_{x,k}(u)
    &:={}
    \E[d_{x,k}(u)\mid\mathcal{H}_k],
    \\
    \Sigma_{d,k}(u)
    &:={}
    \E\!\left[
        \widetilde d_{x,k}(u)\widetilde d_{x,k}(u)^\top
        \mid\mathcal{H}_k
    \right].
\end{aligned}
\label{eq:conditional_prediction_moments}
\end{equation}
For each $\delta_k\in(0,1)$, there exists a compact $\mathcal{H}_k$-measurable conditional quantile region $\mathcal{E}_k(u;\delta_k)\subset\R^n$ satisfying
\begin{equation}
    \Pr\!\left(d_{x,k}(u)\in\mathcal{E}_k(u;\delta_k)
        \mid\mathcal{H}_k\right)
    \geq1-\delta_k
    \quad\text{a.s.}
    \label{eq:conditional_quantile_region}
\end{equation}
The region may depend on the measured state, candidate input, and current conditional law of $\vartheta$.
\end{assumption}
\begin{remark}[Constructing the quantile region]
\label{rem:quantile-construction}
The theory assumes that a conditional quantile region is available. In the implementation, we use a ball whose radius is the conformal order statistic at rank $\lceil(1-\delta_k)(N_{\mathrm{cal}}+1)\rceil$ from $N_{\mathrm{cal}}$ calibration draws of the barrier-channel error norm, where each draw contains a full mission-level realization of $\vartheta$ from its prior. This construction gives marginal $1-\delta_k$ coverage over the calibration distribution, which approximates the $\mathcal{H}_k$-conditional coverage in~\eqref{eq:conditional_quantile_region}. Therefore, the implemented filter approximates the hypothesis of Proposition~\ref{prop:executed-safety}, and we report the empirical violation rates in Section~\ref{sec:sims}. A conditional region, for example one based on online filtering of the law of $\vartheta$, gives the exact hypothesis.
\end{remark}
\subsection{Contraction Theory}
Contraction theory~\cite{lohmiller1998,manchester2017} characterizes incremental stability through the variational dynamics $\delta x_{k+1}=J_k(x_k)\delta x_k$. For $M\succ0$, write $\|z\|_M:=\sqrt{z^\top Mz}$ and let $d_{M_k}(x,y)$ be the corresponding geodesic distance; on a convex Euclidean domain with state-independent $M_k$, $d_{M_k}(x,y)=\|x-y\|_{M_k}$.
\begin{definition}[Discrete contraction metric]
\label{def:metric}
A sequence $\{M_k\}_{k=0}^{N}$ is a discrete contraction metric~\cite{manchester2017,weimccloybao2021} for $F_k$ with factor $\beta\in(0,1)$ if $0<\underline m\leq\overline m<\infty$, $\underline mI\preceq M_k\preceq\overline mI$, and
\begin{equation}
    J_k(x)^\top M_{k+1}J_k(x)\preceq\beta M_k,
    \ \ x\in\mathcal{X},\ 0\leq k<N.
    \label{eq:contraction}
\end{equation}
\end{definition}
The discrete-time linear-quadratic regulator (LQR) Riccati recursion~\cite{weimccloybao2021,weimccloybao2022} provides a trajectory-dependent candidate metric and gain. Given $Q\succ0$, $R\succ0$, and $M_N\succ0$, we compute backward
\begin{equation}
\begin{aligned}
    G_k&:=R+B_k^\top M_{k+1}B_k,
    \\
    K_k&:=G_k^{-1}B_k^\top M_{k+1}A_k,
    \\
    \Phi_k&:=A_k-B_kK_k,
    \\
    M_k&:=Q+A_k^\top M_{k+1}A_k\\
    &\phantom{:=Q}-A_k^\top M_{k+1}B_kG_k^{-1}
      B_k^\top M_{k+1}A_k.
\end{aligned}
\label{eq:riccati}
\end{equation}
Equivalently, $M_k=Q+K_k^\top RK_k+\Phi_k^\top M_{k+1}\Phi_k$, so $M_k\succ0$ by backward induction and
\begin{equation}
    \Phi_k^\top M_{k+1}\Phi_k\prec M_k.
    \label{eq:centerline_contraction}
\end{equation}
Note that this centerline inequality is a synthesis condition, and the exact nonlinear certificate requires~\eqref{eq:contraction} to hold throughout $\mathcal{X}$, which is verified using the following proposition.
\begin{proposition}[Verified differential contraction on $\mathcal{X}$]
\label{prop:tubewise}
Define
\begin{subequations}
\begin{align}
    \beta_k^{\mathrm{ref}}
    &:={}
    \lambda_{\max}\!\left(
        M_k^{-1/2}\Phi_k^\top M_{k+1}\Phi_kM_k^{-1/2}
    \right),
    \label{eq:reference_contraction_factor}
    \\
    \beta_k^{\mathrm{ver}}
    &:={}
    \beta_k^{\mathrm{ref}}
    +\frac{\lambda_{\max}(M_{k+1})}{\lambda_{\min}(M_k)}
      \left(2\|\Phi_k\|_2\Delta_{\Phi,k}+\Delta_{\Phi,k}^2\right).
    \label{eq:global-beta}
\end{align}
\end{subequations}
Let $\beta:=\max_{0\leq k<N}\beta_k^{\mathrm{ver}}$. If $\beta<1$, then
\begin{equation}
    J_k(x)^\top M_{k+1}J_k(x)\preceq\beta M_k,
    \ \ x\in\mathcal{X},\ 0\leq k<N.
    \label{eq:global-differential-contraction}
\end{equation}
Consequently, for states $x,y$ connected by a minimizing geodesic contained in $\mathcal{X}$,
\begin{subequations}
\begin{align}
    d_{M_{k+1}}(F_k(x),F_k(y))
    &\leq\sqrt{\beta}\,d_{M_k}(x,y),
    \label{eq:finite-contraction}
    \\
    d_{M_{k+1}}(F_k(x),\xb_{k+1})
    &\leq\sqrt{\beta}\,d_{M_k}(x,\xb_k).
    \label{eq:reference_contraction_bound}
\end{align}
\end{subequations}
\end{proposition}
\begin{IEEEproof}
Write $J_k(x)=\Phi_k+E_k(x)$ with $\|E_k(x)\|_2\leq\Delta_{\Phi,k}$. By~\eqref{eq:reference_contraction_factor}, $\Phi_k^\top M_{k+1}\Phi_k\preceq\beta_k^{\mathrm{ref}}M_k$, while
\begin{align}
    &\Phi_k^\top M_{k+1}E_k+E_k^\top M_{k+1}\Phi_k+E_k^\top M_{k+1}E_k
    \nonumber\\
    &\quad\preceq
    \frac{\lambda_{\max}(M_{k+1})}{\lambda_{\min}(M_k)}
    \left(2\|\Phi_k\|_2\Delta_{\Phi,k}+\Delta_{\Phi,k}^2\right)M_k.
\end{align}
Adding the bounds gives~\eqref{eq:global-differential-contraction}. Integrating along a minimizing geodesic gives~\eqref{eq:finite-contraction}; setting $y=\xb_k$ gives~\eqref{eq:reference_contraction_bound}.
\end{IEEEproof}
\begin{corollary}[Metric-normalized verified factor]
\label{cor:whitened}
With $s_k:=\bigl(\lambda_{\max}(M_{k+1})/\lambda_{\min}(M_k)\bigr)^{1/2}\Delta_{\Phi,k}$, condition~\eqref{eq:global-differential-contraction} also holds with
\begin{equation}
    \beta_k^{\mathrm{ver}}
    =\beta_k^{\mathrm{ref}}
    +2\sqrt{\beta_k^{\mathrm{ref}}}\,s_k+s_k^2,
    \label{eq:whitened-beta}
\end{equation}
since $\|M_{k+1}^{1/2}E_k(x)M_k^{-1/2}\|_2\leq s_k$ and $\|M_{k+1}^{1/2}\Phi_kM_k^{-1/2}\|_2=\sqrt{\beta_k^{\mathrm{ref}}}$. The factor \eqref{eq:whitened-beta} is sharper than \eqref{eq:global-beta} when the metric is strongly anisotropic, since the eigenvalue ratio enters through its square root.
\end{corollary}
\begin{assumption}[Metric input sensitivity]
\label{ass:input-sensitivity}
The implemented input set satisfies
\begin{equation}
    L_u^M:=\max_{0\leq k<N}
    \sup_{\substack{x\in\mathcal{X},\ u,v\in\mathcal{U}\\u\neq v}}
    \tfrac{\|f(x,u)-f(x,v)\|_{M_{k+1}}}{\|u-v\|_2}<\infty.
\end{equation}
\end{assumption}
This constant quantifies how feedforward sampling and input projection perturb the nominal contracting dynamics; stochastic plant discrepancy enters separately through~\eqref{eq:total_prediction_error}.
\subsection{Stochastic Safety and MPPI Background}
For margins $m_s,m_d>0$, define the baseline margin-adjusted field
\begin{equation}
    h(x,t):=\min\{h_s(x)-m_s,\ h_d(x,t)-m_d\},
    \label{eq:margin_adjusted_safety_field}
\end{equation}
and $\mathcal{S}_k:=\{x\in\mathcal{X}\mid h(x,t_k)\geq0\}$. When every barrier component has one-step input dependence, the analysis uses~\eqref{eq:margin_adjusted_safety_field} directly. For a position-only component of relative degree two, the implementation replaces that component by the lookahead surrogate in Remark~\ref{rem:lookahead}, and $h$ denotes the resulting field. Nested minimum/maximum component fields are allowed, since the executed-input condition uses only the values of $h$.

All the guarantees below are for the selected field $h$ at the discrete sampling instants $t_k$. When $h$ is constructed directly from the physical field, $h(x_k,t_k)\geq0$ implies $h_c(x_k,t_k)\geq\min\{m_s,m_d\}>0$. When a lookahead surrogate is used, the guarantee applies to the surrogate set, and the experiments report physical clearance separately.
\begin{definition}[Time-varying discrete control barrier function]
\label{def:dcbf}
For $\alpha\in(0,1]$, define
\begin{equation}
\begin{aligned}
    \mathcal{K}_h(x,t_k):=\{u\in\mathcal{U}\mid\
    &h(f(x,u),t_{k+1})\\
    &\geq(1-\alpha)h(x,t_k)\}.
\end{aligned}
\end{equation}
The function $h$ is a time-varying DCBF~\cite{agrawal2017,zeng2021} on $\{\mathcal{S}_k\}$ if this set is nonempty for every $x\in\mathcal{S}_k$ and relevant $k$.
\end{definition}
\begin{remark}[Barriers of relative degree two]
\label{rem:lookahead}
For a position-only barrier under Euler discretization, $h(f(x,u),t_{k+1})$ is independent of $u$. Therefore, we use the constant-velocity lookahead barrier $h_{\mathrm{cbf}}(x,t):=h_{\mathrm{phys}}(p+\tau_b(v-v_{\mathrm{obs}}),t)$ with $\tau_b>0$, which depends on the input through the velocity. This barrier is $1$-Lipschitz in $p+\tau_b v$, and we construct the quantile region in the same coordinate. Proposition~\ref{prop:executed-safety} certifies the lookahead set $\{h_{\mathrm{cbf}}\geq0\}$, and we report the physical clearance separately in the simulations.
\end{remark}
Under the nominal dynamics, an input $u_k\in\mathcal{K}_h(x_k,t_k)$ satisfies $h(x_{k+1},t_{k+1})\geq(1-\alpha)h(x_k,t_k)$ and therefore $h(x_k,t_k)\geq(1-\alpha)^k h(x_0,t_0)$. Thus, $\{\mathcal{S}_k\}$ is forward invariant~\cite{agrawal2017} when $x_0\in\mathcal{S}_0$ and the condition holds at every step.
Under stochastic uncertainty, define the quantile-tightened admissible set
\begin{equation}
\begin{aligned}
    \mathcal{K}_{h,k}^{\delta_k}(x_k)
    :=\Bigl\{u\in\mathcal{U}\ \Big|\
    &\inf_{d\in\mathcal{E}_k(u;\delta_k)}
    h(f(x_k,u)+d,t_{k+1})
    \\
    &\geq(1-\alpha)h(x_k,t_k)\Bigr\}.
\end{aligned}
\label{eq:chance_dcbf_admissible_inputs}
\end{equation}
For an $\mathcal{H}_k$-measurable $u_k$ in this set,~\eqref{eq:conditional_quantile_region} implies the one-step condition with conditional probability at least $1-\delta_k$. If $h(x_0,t_0)\geq0$ and the condition is enforced for $k=0,\ldots,N-1$, induction and a union bound give
\begin{equation}
    \Pr\!\left(h(x_k,t_k)\geq0,\ k=0,\ldots,N\right)
    \geq1-\sum_{k=0}^{N-1}\delta_k.
\end{equation}
Thus, when $h(x,t)\geq0$ implies $x\in\mathcal{X}_{\mathrm{safe}}(t)$, the allocation $\sum_k\delta_k\leq\delta_{\mathrm{mis}}$ enforces the mission-level chance constraint in Problem~\ref{prob:main}. With the lookahead construction of Remark~\ref{rem:lookahead}, the same probability bound applies to the surrogate set. We formalize this argument for the input executed by CS-MPC in Proposition~\ref{prop:executed-safety}.
\begin{definition}[MPPI update]
\label{def:mppi}
At MPC step $k$, let $V_k:=(v_{0\mid k},\ldots,v_{H-1\mid k})$. For each rollout, draw $\varepsilon_{i\mid k}^{(b)}\sim\mathcal{N}(0,\Sigma_{i\mid k})$ conditionally independently given $\mathcal{F}_k$, and set $\widetilde u_{i\mid k}^{(b)}:=v_{i\mid k}+\varepsilon_{i\mid k}^{(b)}$ and $u_{i\mid k}^{(b)}:=\Pi_{\mathcal U}(\widetilde u_{i\mid k}^{(b)})$. If $J_k^{(b)}$ is the resulting rollout cost, then
\begin{subequations}
\label{eq:mppi}
\begin{align}
    \omega_k^{(b)}
    &:={}
    \frac{\exp(-(J_k^{(b)}-\rho_k)/\gamma)}
    {\displaystyle\sum_{b'=1}^{N_s}
      \exp(-(J_k^{(b')}-\rho_k)/\gamma)},
    \\
    \rho_k&:=\min_bJ_k^{(b)},
    \\
    u_{\mathrm{MPPI},k}
    &:={}
    \sum_{b=1}^{N_s}\omega_k^{(b)}u_{0\mid k}^{(b)}.
    \label{eq:mppi_first_input}
\end{align}
\end{subequations}
\end{definition}
This importance-weighted update~\cite{williams2017,williams2018information} remains in $\mathcal{U}$ because the set is convex. In the free-energy interpretation inherited from path-integral control~\cite{kappen2005,theodorou2010},
\begin{align*}
    p_k^\star(\widetilde U)
    &=\mathcal Z_k^{-1}e^{-S_k(\widetilde U)/\gamma}
      p_{\mathrm{sample},k}(\widetilde U),
    \\
    p_{\mathrm{sample},k}
    &=\mathcal N\!\left(V_k,
      \operatorname{blkdiag}(\Sigma_{0\mid k},\ldots,\Sigma_{H-1\mid k})\right),
\end{align*}
where $S_k$ is the path cost and $\mathcal Z_k$ is the normalizing constant. The update is a self-normalized importance-sampling estimate that uses only the values of the objective, which allows nonsmooth and nonconvex costs, and it becomes increasingly concentrated on low-cost rollouts as $\gamma$ decreases.
\section{Main Algorithm: The CS-MPC Method}
\label{sec:method}
In this section, we present the two layers of CS-MPC. The \emph{stability and performance layer} uses the contraction-certified tracking feedback inside every sampled rollout, which keeps the rollouts close to the reference. The \emph{execution layer} verifies the selected input using a quantile-tightened DCBF condition before it is applied, which gives the safety guarantee. The rollouts use the nominal model, while the persistent model discrepancy, the process noise, and the actuator implementation error enter the analysis through~\eqref{eq:total_prediction_error}.
\subsection{Contraction Synthesis, Verification, and Stochastic Stability}
\label{ssec:cert}
For a fixed reference, the certificate consists of the metric $\{M_k\}_{k=0}^{N}$, the gain $\{K_k\}_{k=0}^{N-1}$, and a factor $\beta\in(0,1)$ that satisfy~\eqref{eq:contraction} for the exact nonlinear closed-loop Jacobian throughout $\mathcal{X}$. Algorithm~\ref{alg:certificate} computes the metric and the gain along the reference using~\eqref{eq:riccati}, and then verifies the nonlinear inequality using Proposition~\ref{prop:tubewise} or Corollary~\ref{cor:whitened}. The online input projection is treated as an input perturbation in the analysis of Section~\ref{sec:main-results}.
\begin{algorithm}[t]
\caption{Contraction synthesis and domain-wide verification}
\label{alg:certificate}
\begin{algorithmic}[1]
\Require Reference $\{(\xb_k,\ub_k)\}_{k=0}^{N-1}$ and $\xb_N$
\Require Weights $Q\succ0$, $R\succ0$, and terminal metric $M_N\succ0$
\Ensure $\{M_k\}$, $\{K_k\}$, verified factor $\beta<1$, and rate $\lambda$
\For{$k=0,\ldots,N-1$}
    \State $A_k\gets\partial_x f(\xb_k,\ub_k)$ and $B_k\gets\partial_u f(\xb_k,\ub_k)$
\EndFor
\For{$k=N-1,\ldots,0$}
    \State Apply~\eqref{eq:riccati}; set $M_k\gets(M_k+M_k^\top)/2$
\EndFor
\For{$k=0,\ldots,N-1$}
    \State $\Phi_k\gets A_k-B_kK_k$
    \State Compute $\beta_k^{\mathrm{ref}}$ from~\eqref{eq:reference_contraction_factor}
    \State Verify $\Delta_{\Phi,k}\geq\sup_{x\in\mathcal{X}}\|J_k(x)-\Phi_k\|_2$
    \State Compute $\beta_k^{\mathrm{ver}}$ from~\eqref{eq:global-beta} or~\eqref{eq:whitened-beta}
\EndFor
\State $\beta\gets\max_{0\leq k<N}\beta_k^{\mathrm{ver}}$
\If{$\beta\geq1$}
    \State Revise the metric/gain design, certified domain, or Jacobian bound and repeat
\EndIf
\State $\lambda\gets-\log(\beta)/(2\Delta t)$
\State \Return $M$, $K$, $\beta$, and $\lambda$
\end{algorithmic}
\end{algorithm}
The synthesis has complexity $O(N(n^3+n^2m))$. The persistent parameter $\vartheta$, the actuator implementation error, and the process disturbance enter the analysis through $d_{x,k}(u)$. In the following proposition, we give the tracking bounds for the pure-feedback loop.
\begin{proposition}[Nominal and stochastic stability of the contracting feedback]
\label{prop:tube}
Suppose~\eqref{eq:contraction} holds with $\beta<1$ and
\begin{equation}
    \underline mI\preceq M_k\preceq\overline mI,
    \qquad k=0,\ldots,N,
    \label{eq:method_uniform_metric_bounds}
\end{equation}
for $0<\underline m\leq\overline m<\infty$. Assume the state and reference remain in a geodesically convex coordinate chart and define
\begin{equation}
    \mathcal{V}_k:=\|e_k\|_{M_k}^2=e_k^\top M_ke_k.
    \label{eq:method_metric_energy}
\end{equation}
Then:
\emph{(a) Nominal nonlinear stability.} Under $x_{k+1}=F_k(x_k)$,
\begin{equation}
    \mathcal{V}_{k+1}\leq\beta\mathcal{V}_k,
    \qquad
    \|e_k\|_2\leq
    \sqrt{\frac{\overline m}{\underline m}}\,\beta^{k/2}\|e_0\|_2.
\end{equation}
\emph{(b) Stochastic prediction error.} Consider $x_{k+1}=F_k(x_k)+d_{x,k}(u_k^{\mathrm{fb}}(x_k))$ and assume $u_k^{\mathrm{fb}}(x_k)\in\mathcal{U}$. Let $\bar d_k^{\mathrm{fb}}:=\bar d_{x,k}(u_k^{\mathrm{fb}}(x_k))$ and $\Sigma_{d,k}^{\mathrm{fb}}:=\Sigma_{d,k}(u_k^{\mathrm{fb}}(x_k))$. Choose $\eta_d>0$ so that $\beta_d:=(1+\eta_d)\beta<1$, and define
\[
    c_{d,k}:=(1+\eta_d^{-1})\|\bar d_k^{\mathrm{fb}}\|_{M_{k+1}}^2
    +\operatorname{tr}(M_{k+1}\Sigma_{d,k}^{\mathrm{fb}}).
\]
Then
\begin{equation}
    \E[\mathcal{V}_{k+1}\mid\mathcal{H}_k]
    \leq\beta_d\mathcal{V}_k+c_{d,k}.
    \label{eq:conditional_metric_recursion}
\end{equation}
If $c_{d,k}\leq\overline c_d$ almost surely, then
\begin{equation}
    \E[\mathcal{V}_k]
    \leq\beta_d^k\E[\mathcal{V}_0]
    +\overline c_d\frac{1-\beta_d^k}{1-\beta_d}.
    \label{eq:nltube}
\end{equation}
If $\bar d_k^{\mathrm{fb}}=0$, the sharper recursion is
\begin{equation}
    \E[\mathcal{V}_{k+1}\mid\mathcal{H}_k]
    \leq\beta\mathcal{V}_k
    +\operatorname{tr}(M_{k+1}\Sigma_{d,k}^{\mathrm{fb}}),
    \label{eq:centered_stochastic_metric_recursion}
\end{equation}
which is the discrete-time mean-square stochastic contraction relation~\cite{tsukamoto2021}.
\end{proposition}
\begin{IEEEproof}
Part~(a) follows from~\eqref{eq:reference_contraction_bound} and~\eqref{eq:method_uniform_metric_bounds}. For part~(b), set $g_k:=F_k(x_k)\ominus F_k(\xb_k)$, so $e_{k+1}=g_k+d_{x,k}(u_k^{\mathrm{fb}}(x_k))$. Conditioning on $\mathcal{H}_k$ gives
\begin{align}
    \E[\mathcal{V}_{k+1}\mid\mathcal{H}_k]
    ={}&\|g_k\|_{M_{k+1}}^2
    +2g_k^\top M_{k+1}\bar d_k^{\mathrm{fb}}
    \nonumber\\
    &+\|\bar d_k^{\mathrm{fb}}\|_{M_{k+1}}^2
    +\operatorname{tr}(M_{k+1}\Sigma_{d,k}^{\mathrm{fb}}).
\end{align}
Young's inequality bounds the cross term by $\eta_d\|g_k\|_{M_{k+1}}^2+\eta_d^{-1}\|\bar d_k^{\mathrm{fb}}\|_{M_{k+1}}^2$. Combining this with $\|g_k\|_{M_{k+1}}^2\leq\beta\mathcal{V}_k$ proves~\eqref{eq:conditional_metric_recursion}; expectation and iteration give~\eqref{eq:nltube}, and centering removes the cross term.
\end{IEEEproof}
\begin{remark}[Interpretation]
The conditional mean is the bias due to the mission parameter and the actuator implementation error, and the covariance is due to the remaining model uncertainty and the process noise. Note that this mean-square recursion is separate from the one-step quantile region used by the execution layer.
\end{remark}
\subsection{Contractive Sampling Law}
\label{ssec:sampling}
At step $k$, rollout $b$ starts from $x_{0\mid k}^{(b)}=x_k$ and, for $i=0,\ldots,H-1$, uses
\begin{subequations}
\label{eq:csampling}
\begin{align}
    \widetilde u_{i\mid k}^{(b)}
    &:={}
    \ub_{k+i}+K_{k+i}(\xb_{k+i}\ominus x_{i\mid k}^{(b)})
    +\mu_{i\mid k}+\varepsilon_{i\mid k}^{(b)},
    \label{eq:csampling_unprojected}
    \\
    u_{i\mid k}^{(b)}
    &:={}
    \Pi_{\mathcal U}(\widetilde u_{i\mid k}^{(b)}),
    \label{eq:csampling_projected}
    \\
    x_{i+1\mid k}^{(b)}
    &:={}
    f(x_{i\mid k}^{(b)},u_{i\mid k}^{(b)}).
\end{align}
\end{subequations}
Conditioned on $\mathcal F_k$, the free perturbations are independent and satisfy $\varepsilon_{i\mid k}^{(b)}\sim\mathcal N(0,\Sigma_{i\mid k})$. Each sample set also includes three deterministic anchors (the mean, pure-feedback, and backup rollouts), where the backup is available to the safety filter but is excluded from the MPPI average. With $\mathcal B_k$ denoting the eligible non-backup set and $\Delta\mu_{i\mid k}:=\sum_{b\in\mathcal B_k}\omega_k^{(b)}\varepsilon_{i\mid k}^{(b)}$, the warm start is
\begin{equation}
    \mu_{\cdot\mid k+1}
    =\operatorname{shift}\!\left(\lambda_\mu
        (\mu_{\cdot\mid k}+\Delta\mu_{\cdot\mid k})\right),
    \label{eq:warmstart}
\end{equation}
where $\lambda_\mu\in(0,1]$ and the experiments use $\lambda_\mu=0.95$.
\subsubsection{Barrier-Aware Exploration Covariance}
For an integer warning horizon $H_w\in\{0,\ldots,H\}$ and a barrier threshold $h_{\mathrm{warn}}>0$, we widen isotropic exploration when the predicted barrier margin is small. Let
\begin{equation}
    \underline h_k:=\min_{0\leq i\leq H_w}\widehat h_{i\mid k}^{\mathrm{prev}},
    \label{eq:predicted_warning_margin}
\end{equation}
where the predicted barrier values are shifted from the previously accepted solution; set $\underline h_0:=h(x_0,t_0)$. With nominal sampling standard deviation $\sigma>0$ and inflation factor $\zeta\geq1$, the isotropic rule is
\begin{equation}
    \Sigma_{i\mid k}=\begin{cases}
        \zeta^2\sigma^2I,&\underline h_k<h_{\mathrm{warn}},\\
        \sigma^2I,&\underline h_k\geq h_{\mathrm{warn}},
    \end{cases}
    \quad i=0,\ldots,H-1.
    \label{eq:widen}
\end{equation}
Alternatively, let $R_{i\mid k}^M:=B_{k+i}^\top M_{k+i+1}B_{k+i}$ and choose a metric-shaped exploration scale $s_{i\mid k}\geq0$. When $R_{i\mid k}^M\succ0$, set
\begin{equation}
    \Sigma_{i\mid k}=s_{i\mid k}^2(R_{i\mid k}^M)^{-1},
    \label{eq:budgetcov}
\end{equation}
which gives $\operatorname{tr}(B_{k+i}^\top M_{k+i+1}B_{k+i}\Sigma_{i\mid k})=ms_{i\mid k}^2$. A regularized inverse is used when $R_{i\mid k}^M$ is singular or ill-conditioned.

If $T_p$ selects position coordinates and $h(\cdot,t_k)$ is $L_h$-Lipschitz on the relevant reachable set, define
\begin{subequations}
\begin{align}
    h_{\mathrm{eff},k}
    &:={}
    \min\{h(x_k,t_k),\underline h_k\},
    \label{eq:effective_barrier_margin}
    \\
    \mathcal V_{\mathrm{allow},k}
    &:={}
    \frac{([h_{\mathrm{eff},k}]_+/L_h)^2}
    {\lambda_{\max}(T_pM_k^{-1}T_p^\top)}.
\end{align}
\end{subequations}
With nominal scale $s_{\mathrm{nom}}>0$, we use the scalar cap
\begin{equation}
    s_{0\mid k}^2
    =\min\!\left\{\frac{(1-\beta)\mathcal V_{\mathrm{allow},k}}{m},
        \zeta^2s_{\mathrm{nom}}^2\right\}.
    \label{eq:budget}
\end{equation}
The cap may instead be applied only along an input-space barrier-normal direction, which leaves the tangential exploration widened. This covariance design is an exploration heuristic, and the safety of the executed input is verified separately by the execution layer.
In the following proposition, we prove that the spread of the feedback-closed rollouts is bounded uniformly in the horizon, which is the mechanism behind the sample efficiency observed in Section~\ref{sec:sims}.
\begin{proposition}[Horizon-uniform rollout concentration]
\label{prop:concentration}
Suppose that the contraction condition~\eqref{eq:global-differential-contraction} holds with $\beta<1$ on a convex domain $\mathcal X$ in a Euclidean chart and that Assumption~\ref{ass:input-sensitivity} holds. Let $x_{i\mid k}^{\mathrm{fb}}$ denote the pure-feedback anchor rollout, which is~\eqref{eq:csampling} with $\mu_{i\mid k}=0$ and $\varepsilon_{i\mid k}=0$, and suppose that $x_{i\mid k}^{(b)},x_{i\mid k}^{\mathrm{fb}}\in\mathcal X$ and $u_{k+i}^{\mathrm{fb}}(x_{i\mid k}^{(b)})\in\mathcal U$ for $i=0,\ldots,H-1$. Define the rollout spread $\varsigma_{i\mid k}^{(b)}:=\|x_{i\mid k}^{(b)}-x_{i\mid k}^{\mathrm{fb}}\|_{M_{k+i}}$. Then $\varsigma_{0\mid k}^{(b)}=0$ and
\begin{equation}
    \varsigma_{i+1\mid k}^{(b)}
    \leq\sqrt\beta\,\varsigma_{i\mid k}^{(b)}
    +L_u^M\|\mu_{i\mid k}+\varepsilon_{i\mid k}^{(b)}\|_2.
    \label{eq:spread_recursion}
\end{equation}
If, in addition, $\|\mu_{i\mid k}\|_2\leq\bar\mu$ and $\operatorname{tr}(\Sigma_{i\mid k})\leq\bar\sigma^2$ for $i=0,\ldots,H-1$, then for every $i\in\{0,\ldots,H\}$,
\begin{equation}
    \left(\E\bigl[(\varsigma_{i\mid k}^{(b)})^2\mid\mathcal F_k\bigr]\right)^{1/2}
    \leq\frac{L_u^M\sqrt{\bar\mu^2+\bar\sigma^2}}{1-\sqrt\beta},
    \label{eq:spread_bound}
\end{equation}
which is uniform in the horizon $H$.
\end{proposition}
\begin{IEEEproof}
Using the closed-loop map~\eqref{eq:closed_loop_map}, the anchor rollout satisfies $x_{i+1\mid k}^{\mathrm{fb}}=F_{k+i}(x_{i\mid k}^{\mathrm{fb}})$, so that
\begin{equation*}
\begin{aligned}
    x_{i+1\mid k}^{(b)}-x_{i+1\mid k}^{\mathrm{fb}}
    ={}&f(x_{i\mid k}^{(b)},u_{i\mid k}^{(b)})-F_{k+i}(x_{i\mid k}^{(b)})\\
    &+F_{k+i}(x_{i\mid k}^{(b)})-F_{k+i}(x_{i\mid k}^{\mathrm{fb}}).
\end{aligned}
\end{equation*}
Since $u_{k+i}^{\mathrm{fb}}(x_{i\mid k}^{(b)})\in\mathcal U$ and the projection onto the convex set $\mathcal U$ is nonexpansive, we have $\|u_{i\mid k}^{(b)}-u_{k+i}^{\mathrm{fb}}(x_{i\mid k}^{(b)})\|_2\leq\|\mu_{i\mid k}+\varepsilon_{i\mid k}^{(b)}\|_2$. Therefore, Assumption~\ref{ass:input-sensitivity} bounds the first difference by $L_u^M\|\mu_{i\mid k}+\varepsilon_{i\mid k}^{(b)}\|_2$, and~\eqref{eq:finite-contraction} bounds the second difference by $\sqrt\beta\,\varsigma_{i\mid k}^{(b)}$, which proves~\eqref{eq:spread_recursion}. Iterating~\eqref{eq:spread_recursion} from $\varsigma_{0\mid k}^{(b)}=0$ gives $\varsigma_{i\mid k}^{(b)}\leq L_u^M\sum_{j=0}^{i-1}\beta^{(i-1-j)/2}\|\mu_{j\mid k}+\varepsilon_{j\mid k}^{(b)}\|_2$. Because $\mu_{j\mid k}$ is $\mathcal F_k$-measurable and $\varepsilon_{j\mid k}^{(b)}$ is conditionally zero-mean with covariance $\Sigma_{j\mid k}$, we have $\E[\|\mu_{j\mid k}+\varepsilon_{j\mid k}^{(b)}\|_2^2\mid\mathcal F_k]=\|\mu_{j\mid k}\|_2^2+\operatorname{tr}(\Sigma_{j\mid k})\leq\bar\mu^2+\bar\sigma^2$. Using Minkowski's inequality and the geometric sum, we obtain~\eqref{eq:spread_bound}.
\end{IEEEproof}
\begin{remark}[Open-loop rollouts and scope]
For open-loop rollouts, the same argument gives~\eqref{eq:spread_recursion} with $\sqrt\beta$ replaced by a Lipschitz constant of $f(\cdot,u)$ in the metric, which is larger than one for unstable dynamics, so that the bound on the spread grows geometrically with $H$. Note that Proposition~\ref{prop:concentration} bounds the dispersion of the rollouts, and Theorem~\ref{thm:tracking} gives the finite-sample bound for the implemented controller.
\end{remark}
\subsection{Rollout Cost and Terminal Contraction Cost}
\label{ssec:cost}
With $e_{i\mid k}^{(b)}:=x_{i\mid k}^{(b)}\ominus\xb_{k+i}$, the rollout objective is
\begin{align}
    J_k^{(b)}
    &:={}
    \sum_{i=0}^{H-1}\!\left(
        \|e_{i\mid k}^{(b)}\|_{M_{k+i}}^2
        +c_s(x_{i\mid k}^{(b)},u_{i\mid k}^{(b)})
        +\rho_\varepsilon\|\varepsilon_{i\mid k}^{(b)}\|_2^2
    \right)
    \nonumber\\
    &\quad+\rho_f\|e_{H\mid k}^{(b)}\|_{M_{k+H}}^2
    +J_{\mathrm{DCBF},k}^{(b)},
    \label{eq:cost}
\end{align}
where $\rho_\varepsilon\geq0$, $\rho_f\geq1$, and the DCBF term guides the proposal distribution toward the safe rollouts.

The stage cost $c_s(x,u)\geq0$ can be nonsmooth or discontinuous, because it enters only through the weights in~\eqref{eq:mppi}. In the simulations, we use the static-violation penalties $c_s\supseteq w_s[m_s^{\mathrm{cost}}-h_s(x)]_+^2+w_{\mathrm{hard}}\mathbf 1_{\{h_s(x)<0\}}$, along with indicator and minimum-impulse $\ell_0$ terms. These terms and $J_{\mathrm{DCBF},k}^{(b)}$ are nonnegative, so that $J_k^{(b)}\geq0$, which is required by the Hoeffding bounds on the weighted input.
\subsection{Executed-Input DCBF Filter and One-Step Chance Tightening}
\label{ssec:filter}
The rollout DCBF penalties guide the search, while the nonlinear DCBF-admissible input set can be nonconvex, so that the weighted average of admissible inputs can lie outside this set. Therefore, CS-MPC verifies the input at execution using the DCBF condition. For a candidate $u\in\mathcal U$, write $\widehat x_{k+1}(u):=f(x_k,u)$ and $x_{k+1}=\widehat x_{k+1}(u)+d_{x,k}(u)$. Let $\widetilde{\mathcal E}_k(u;\delta_k):=\mathcal E_k(u;\delta_k)\cup\{0\}$ and define
\begin{equation}
\begin{aligned}
    \Delta_{h,k}^{\delta_k}(u)
    :=\sup_{d\in\widetilde{\mathcal E}_k(u;\delta_k)}\bigl[
    &h(\widehat x_{k+1}(u),t_{k+1})
    \\
    &-h(\widehat x_{k+1}(u)+d,t_{k+1})\bigr].
\end{aligned}
\label{eq:exact-barrier-tightening}
\end{equation}
Because $0$ belongs to the augmented region, $\Delta_{h,k}^{\delta_k}(u)\geq0$. The executed-input DCBF condition is
\begin{equation}
\begin{aligned}
    \mathcal C_k^{\delta_k}(u):\quad
    &h(\widehat x_{k+1}(u),t_{k+1})
    -\Delta_{h,k}^{\delta_k}(u)\\
    &\geq(1-\alpha)h(x_k,t_k).
\end{aligned}
    \label{eq:executed-dcbf-test}
\end{equation}
Equivalently, the infimum of $h(f(x_k,u)+d,t_{k+1})$ over the augmented quantile region must exceed $(1-\alpha)h(x_k,t_k)$. On the event $d_{x,k}(u)\in\mathcal E_k(u;\delta_k)$,
\begin{equation}
    h(x_{k+1},t_{k+1})
    \geq h(\widehat x_{k+1}(u),t_{k+1})
    -\Delta_{h,k}^{\delta_k}(u).
    \label{eq:tightening_implication}
\end{equation}
This is the execution-time counterpart of quantile constraint tightening in stochastic predictive control~\cite{mesbah2016,nakka2023chance}.

A sufficient norm-based specialization uses $r_{x,k}^{\delta_k}(u):=\sup_{d\in\widetilde{\mathcal E}_k(u;\delta_k)}\|d\|_2$. If $h(\cdot,t_{k+1})$ is $L_h$-Lipschitz, then $\Delta_{h,k}^{\delta_k}(u)\leq L_hr_{x,k}^{\delta_k}(u)$, so it suffices that
\begin{equation*}
    h(\widehat x_{k+1}(u),t_{k+1})-L_hr_{x,k}^{\delta_k}(u)
    \geq(1-\alpha)h(x_k,t_k).
\end{equation*}
Direct evaluation of~\eqref{eq:exact-barrier-tightening} can be less conservative.
\begin{assumption}[Chance-certified backup policy]
\label{ass:backup}
For each $\delta_k$, a backup $\kappa_B(x,t_k)\in\mathcal U$ is available on the trigger region and satisfies $\mathcal C_k^{\delta_k}(\kappa_B(x,t_k))$ under the same conditional quantile model used for the performance candidates.
\end{assumption}
Assumption~\ref{ass:backup} states that a certified action exists whenever the filter reaches the backup branch. 
\begin{remark}
    The constructive backups used in the simulation studies are clearance-gradient evasion for the rover, thrust-tilt evasion for the quadrotor, and radial retreat for the asteroid. If no candidate, including the backup, is certified, then Algorithm~\ref{alg:csmc} flags the step and applies the candidate with the largest predicted barrier value. In the implementation, a candidate is certified when it satisfies~\eqref{eq:executed-dcbf-test} and the recoverability check described in Section~\ref{sec:sims}. Note that Proposition~\ref{prop:executed-safety} is conditional on the backup assumption.
\end{remark}
In the following proposition, we prove the mission-level probabilistic safety of the input applied by Algorithm~\ref{alg:csmc}.
\begin{proposition}[Mission-level probabilistic safety of the executed input]
\label{prop:executed-safety}
Suppose $h(x_0,t_0)\geq0$, Assumptions~\ref{ass:stochastic-prediction-error} and~\ref{ass:backup} hold, and
\begin{equation}
    \sum_{k=0}^{N-1}\delta_k\leq\delta_{\mathrm{mis}}.
    \label{eq:method_risk_allocation}
\end{equation}
If Algorithm~\ref{alg:csmc} applies an input satisfying~\eqref{eq:executed-dcbf-test} at every step, then
\begin{equation}
\begin{aligned}
    \Pr\!\left(h(x_k,t_k)\geq0,\ k=0,\ldots,N\right)
    &\geq1-\sum_{k=0}^{N-1}\delta_k\\
    &\geq1-\delta_{\mathrm{mis}}.
\end{aligned}
    \label{eq:executed_mission_safety}
\end{equation}
\end{proposition}
\begin{IEEEproof}
The selected input $u_k^\star$ is $\mathcal H_k$-measurable. Let $\mathsf E_k:=\{d_{x,k}(u_k^\star)\in\mathcal E_k(u_k^\star;\delta_k)\}$. By~\eqref{eq:conditional_quantile_region}, $\Pr(\mathsf E_k\mid\mathcal H_k)\geq1-\delta_k$. On $\mathsf E_k$, equations~\eqref{eq:executed-dcbf-test} and~\eqref{eq:tightening_implication} imply $h(x_{k+1},t_{k+1})\geq(1-\alpha)h(x_k,t_k)$. Induction and a union bound yield~\eqref{eq:executed_mission_safety}.
\end{IEEEproof}
\subsection{Online CS-MPC Procedure}
\label{ssec:online_algorithm}
Algorithm~\ref{alg:csmc} summarizes one receding-horizon update, which has three stages. In Stage~1, we set the exploration covariance using the predicted barrier margin, and we draw the free perturbations along with the mean, feedback, and backup anchors. In Stage~2, we propagate the feedback-closed rollouts using~\eqref{eq:csampling}, evaluate the cost~\eqref{eq:cost}, and compute the weighted input over the eligible set. In Stage~3, the execution layer verifies the weighted input using~\eqref{eq:executed-dcbf-test}, and when the weighted input does not satisfy it, the execution layer applies the candidate that satisfies this condition and is closest to the weighted input, or the backup input. Every candidate is projected onto $\mathcal U$ before its cost is computed and before it is verified. The weighted average remains in $\mathcal U$ because $\mathcal U$ is convex, but it must still satisfy the nonlinear quantile-tightened DCBF condition.
\begin{algorithm}[t]
\caption{CS-MPC update with executed-input probabilistic safety verification}
\label{alg:csmc}
\begin{algorithmic}[1]
\Require Measured state $x_k$ and reference/certificate window
\Require Warm-start mean $\mu_{\cdot\mid k}$ and previous predicted barrier values
\Require Barrier $h$, risk $\delta_k$, quantile/tightening evaluator, and backup $\kappa_B$
\Ensure Applied input $u_k^\star$ and shifted mean $\mu_{\cdot\mid k+1}$
\State Compute $\underline h_k$ from~\eqref{eq:predicted_warning_margin}
\State Compute $h_{\mathrm{eff},k}$ from~\eqref{eq:effective_barrier_margin}
\If{isotropic exploration is used}
    \State Set $\Sigma_{i\mid k}$ using~\eqref{eq:widen}
\Else
    \State Set the metric-shaped covariance using~\eqref{eq:budgetcov}--\eqref{eq:budget}
\EndIf
\State Draw the free perturbations and insert the mean, feedback, and backup anchors
\For{$b=1,\ldots,N_s$}
    \State $x_{0\mid k}^{(b)}\gets x_k$
    \For{$i=0,\ldots,H-1$}
        \State Compute $u_{i\mid k}^{(b)}$ from~\eqref{eq:csampling_unprojected}--\eqref{eq:csampling_projected}
        \State $x_{i+1\mid k}^{(b)}\gets f(x_{i\mid k}^{(b)},u_{i\mid k}^{(b)})$
    \EndFor
    \State Evaluate $J_k^{(b)}$ from~\eqref{eq:cost}
\EndFor
\State Compute weights over the eligible non-backup set $\mathcal B_k$
\State $u_k^{\mathrm{avg}}\gets\sum_{b\in\mathcal B_k}\omega_k^{(b)}u_{0\mid k}^{(b)}$
\If{$\mathcal C_k^{\delta_k}(u_k^{\mathrm{avg}})$ holds}
    \State $u_k^\star\gets u_k^{\mathrm{avg}}$
\Else
    \State Verify all sampled first inputs at $x_k$ using $\mathcal C_k^{\delta_k}$
    \If{at least one sampled input satisfies $\mathcal C_k^{\delta_k}$}
        \State $u_k^\star\gets$ candidate that satisfies $\mathcal C_k^{\delta_k}$ and is closest to $u_k^{\mathrm{avg}}$
    \Else
        \State $u_k^\star\gets\kappa_B(x_k,t_k)$ if it satisfies $\mathcal C_k^{\delta_k}$; otherwise, flag the step and apply the candidate with the largest predicted $h$
    \EndIf
\EndIf
\State Update $\mu_{\cdot\mid k+1}$ using~\eqref{eq:warmstart}
\State \Return $u_k^\star$ and $\mu_{\cdot\mid k+1}$
\end{algorithmic}
\end{algorithm}
\noindent\textbf{Input saturation:} All sampled and executed inputs are projected onto $\mathcal U$, and the DCBF condition is verified after the projection. The projection changes the input relative to the unconstrained feedback, and this change is included as an input perturbation in Theorem~\ref{thm:tracking}. Therefore, the nominal contraction certificate and the safety verification apply to the projected input.
\section{Finite-Sample Stochastic Tracking and Executed-Input Safety}
\label{sec:main-results}
In this section, we derive the main results for the implemented controller, which uses a finite number of samples. We first bound the error between the weighted control input and its population value (Lemma~\ref{lem:csmc-concentration}). We then propagate this error, along with the plant-prediction error, through the contracting dynamics to obtain the stability bound (Theorem~\ref{thm:tracking}). The safety result for the executed input is given separately by Proposition~\ref{prop:executed-safety}. Because the sample set contains three deterministic anchors, $N_f:=N_s-3\geq1$ free rollouts remain.
\subsection{Concentration of the Weighted CS-MPC Input}
\label{ssec:weighted-concentration}
Fix step $k$ and condition on $\mathcal F_k$. For free rollout $b$, write $u_k^{(b)}:=u_{0\mid k}^{(b)}$ and $q_k^{(b)}:=\exp(-J_k^{(b)}/\gamma)$. Let $\mathcal D:=\{\mathrm{mean},\mathrm{feedback}\}$ denote the deterministic anchors included in the average, and define $a_k:=\sum_{d\in\mathcal D}q_k^{(d)}u_k^{(d)}$ and $z_k:=\sum_{d\in\mathcal D}q_k^{(d)}$. The weighted control input over the eligible set is
\begin{equation}
    \widehat u_k^{(N_f)}
    :=\frac{a_k+\sum_{b=1}^{N_f}q_k^{(b)}u_k^{(b)}}
    {z_k+\sum_{b=1}^{N_f}q_k^{(b)}}.
    \label{eq:finite-csmc-input}
\end{equation}
Minimum-cost subtraction before exponentiation leaves this ratio unchanged. For a representative free rollout, let $Y_k:=\E[q_k^{(1)}u_k^{(1)}\mid\mathcal F_k]$ and $Z_k:=\E[q_k^{(1)}\mid\mathcal F_k]$; the corresponding conditional finite-batch population input is $u_k^{\circ,(N_f)}:=(a_k+N_fY_k)/(z_k+N_fZ_k)$.
\begin{assumption}[Weighted-input denominator]
\label{ass:weighted-moments}
There exists $q_{\min}>0$ such that $Z_k\geq q_{\min}$ for every relevant step and almost every reachable pre-sampling information state.
\end{assumption}
Because $J_k^{(b)}\geq0$, $0<q_k^{(b)}\leq1$. Compactness of $\mathcal U$ gives $u_{\max}:=\max_{u\in\mathcal U}\|u\|_2<\infty$ and $d_{\mathcal U}:=\operatorname{diam}(\mathcal U)<\infty$. A uniform bound $J_k^{(b)}\leq J_{\max}$ is sufficient for Assumption~\ref{ass:weighted-moments}, with $q_{\min}=e^{-J_{\max}/\gamma}$. Hard penalties can make $q_{\min}$ small, whereas bounded indicator and $\ell_0$ terms keep the bound meaningful. Define
\begin{equation}
\begin{aligned}
    C_u&:=\frac{2u_{\max}}{q_{\min}}
    \left(\sqrt{2m}+\frac{1}{\sqrt2}\right),
    \\
    \varepsilon_u(N_f,\eta_s)
    &:=C_u\sqrt{\frac{\log(4m/\eta_s)}{N_f}}.
\end{aligned}
\label{eq:concentration-constant}
\end{equation}
In the following lemma, we bound the error between the weighted control input and its population value, which will be used in Theorem~\ref{thm:tracking}.
\begin{lemma}[Finite-sample concentration of the weighted CS-MPC input]
\label{lem:csmc-concentration}
Under Assumption~\ref{ass:weighted-moments}, fix $\eta_s\in(0,1)$ and suppose
\begin{equation}
    N_f\geq N_0(\eta_s):=
    \left\lceil\frac{2}{q_{\min}^2}\log\!\left(\frac{4}{\eta_s}\right)\right\rceil.
    \label{eq:minimum-concentration-budget}
\end{equation}
Then
\begin{equation}
    \Pr\!\left(
        \|\widehat u_k^{(N_f)}-u_k^{\circ,(N_f)}\|_2
        \leq\varepsilon_u(N_f,\eta_s)
        \mid\mathcal F_k
    \right)\geq1-\eta_s.
    \label{eq:csmc-concentration}
\end{equation}
\end{lemma}
\begin{IEEEproof}
All the statements below are conditional on $\mathcal F_k$, for which $a_k$, $z_k$, $Y_k$, and $Z_k$ are deterministic and the pairs $(q_k^{(b)},u_k^{(b)})$, $b=1,\ldots,N_f$, are independent and identically distributed, because the free perturbations are independent across the rollouts. Let $\widehat Y_k:=N_f^{-1}\sum_bq_k^{(b)}u_k^{(b)}$, $\widehat Z_k:=N_f^{-1}\sum_bq_k^{(b)}$, $\bar a_k:=a_k/N_f$, and $\bar z_k:=z_k/N_f\geq0$, so that
\begin{equation}
    \widehat u_k^{(N_f)}=\frac{\bar a_k+\widehat Y_k}{\bar z_k+\widehat Z_k},
    \qquad
    u_k^{\circ,(N_f)}=\frac{\bar a_k+Y_k}{\bar z_k+Z_k}.
    \label{eq:ratio-form}
\end{equation}
Since $q_k^{(b)}\in(0,1]$ and each component of $q_k^{(b)}u_k^{(b)}$ lies in $[-u_{\max},u_{\max}]$, Hoeffding's inequality~\cite{hoeffding1963} gives $\Pr(|\widehat Z_k-Z_k|\geq t\mid\mathcal F_k)\leq2\exp(-2N_ft^2)$ and, for each component $j$, $\Pr(|\widehat Y_{k,j}-Y_{k,j}|\geq s\mid\mathcal F_k)\leq2\exp(-N_fs^2/(2u_{\max}^2))$. Setting these bounds equal to $\eta_s/2$ and $\eta_s/(2m)$, and using a union bound along with $\|\widehat Y_k-Y_k\|_2\leq\sqrt m\max_j|\widehat Y_{k,j}-Y_{k,j}|$, we obtain, with probability at least $1-\eta_s$,
\begin{align}
    |\widehat Z_k-Z_k|
    &\leq\sqrt{\frac{\log(4/\eta_s)}{2N_f}},
    \nonumber\\
    \|\widehat Y_k-Y_k\|_2
    &\leq u_{\max}\sqrt{\frac{2m\log(4m/\eta_s)}{N_f}}.
    \label{eq:hoeffding-events}
\end{align}
On this event, the budget~\eqref{eq:minimum-concentration-budget} gives $|\widehat Z_k-Z_k|\leq q_{\min}/2$, so that $\bar z_k+\widehat Z_k\geq Z_k-|\widehat Z_k-Z_k|\geq q_{\min}/2$ by Assumption~\ref{ass:weighted-moments}. Since every anchor input and every sampled input belongs to $\mathcal U$, we have $\|\bar a_k+Y_k\|_2\leq u_{\max}(\bar z_k+Z_k)$, which gives $\|u_k^{\circ,(N_f)}\|_2\leq u_{\max}$. Subtracting the two expressions in~\eqref{eq:ratio-form}, we obtain
\begin{equation}
    \widehat u_k^{(N_f)}-u_k^{\circ,(N_f)}
    =\frac{(\widehat Y_k-Y_k)-u_k^{\circ,(N_f)}(\widehat Z_k-Z_k)}{\bar z_k+\widehat Z_k},
    \label{eq:ratio-identity}
\end{equation}
and therefore
\begin{equation}
\begin{aligned}
    \|\widehat u_k^{(N_f)}-u_k^{\circ,(N_f)}\|_2
    \leq\frac{2}{q_{\min}}\bigl(
    &\|\widehat Y_k-Y_k\|_2
    \\
    &+u_{\max}|\widehat Z_k-Z_k|\bigr).
\end{aligned}
\label{eq:ratio-concentration-step}
\end{equation}
Substituting~\eqref{eq:hoeffding-events} in~\eqref{eq:ratio-concentration-step} and using $\log(4/\eta_s)\leq\log(4m/\eta_s)$ gives the bound $\varepsilon_u(N_f,\eta_s)$ in~\eqref{eq:concentration-constant}, which proves~\eqref{eq:csmc-concentration}.
\end{IEEEproof}
Both inputs in~\eqref{eq:csmc-concentration} belong to the convex set $\mathcal U$, so that their difference is bounded by $d_{\mathcal U}$. Splitting the conditional second moment over the concentration event of Lemma~\ref{lem:csmc-concentration} and its complement gives
\begin{equation}
\begin{aligned}
    \E\!\left[\|\widehat u_k^{(N_f)}-u_k^{\circ,(N_f)}\|_2^2
        \mid\mathcal F_k\right]
    &\leq\varepsilon_{\mathrm{samp}}^2(N_f,\eta_s),
    \\
    \varepsilon_{\mathrm{samp}}^2(N_f,\eta_s)
    &:=C_u^2\tfrac{\log(4m/\eta_s)}{N_f}
      +\eta_sd_{\mathcal U}^2.
\end{aligned}
\label{eq:conditional_sampling_mse}
\end{equation}
For example, $\eta_s=N_f^{-1}$ gives $\varepsilon_{\mathrm{samp}}^2=O((\log N_f)/N_f)$ whenever~\eqref{eq:minimum-concentration-budget} holds.
The constant $C_u$ scales with $u_{\max}/q_{\min}$, so Lemma~\ref{lem:csmc-concentration} is a consistency result with the rate $N_f^{-1/2}$, and its constant is conservative when the sampled inputs are spread over $\mathcal U$. The bound is tightened when the sampling is restricted to a neighborhood of the feedback input, as stated in the following corollary.
\begin{corollary}[Shifted concentration for bounded residual sampling]
\label{cor:shifted-concentration}
Let $v_k$ be $\mathcal F_k$-measurable. If every eligible first input satisfies $\|u_k^{(b)}-v_k\|_2\leq r_u$, then Lemma~\ref{lem:csmc-concentration} and~\eqref{eq:conditional_sampling_mse} hold with $u_{\max}$ replaced by $r_u$ and $d_{\mathcal U}$ replaced by $2r_u$. In addition, the second-moment bound in~\eqref{eq:conditional_sampling_mse} is at most $4r_u^2$.
\end{corollary}
\begin{IEEEproof}
It is sufficient to subtract $v_k$ from both ratios in~\eqref{eq:finite-csmc-input}, which leaves their difference unchanged. The proof of Lemma~\ref{lem:csmc-concentration} then applies to $q_k^{(b)}(u_k^{(b)}-v_k)$, and both shifted ratios lie in the ball of radius $r_u$, which gives the cap $4r_u^2$.
\end{IEEEproof}
One implementation uses $v_k=u_k^{\mathrm{fb}}$ and clips each feedforward residual to a Euclidean ball of radius $r_u$ before the projection onto $\mathcal U$. Since the projection is nonexpansive, $\|\Pi_{\mathcal U}(u_k^{\mathrm{fb}}+a)-\Pi_{\mathcal U}(u_k^{\mathrm{fb}})\|_2\leq\|a\|_2\leq r_u$, which gives the radius $r_u$ analytically. With this residual clipping and the feedback input as the backup, every sampled, weighted, and backup input is within $r_u$ of $u_k^{\mathrm{fb}}$ whenever $u_k^{\mathrm{fb}}\in\mathcal U$, so that Assumptions~\ref{ass:population-deviation} and~\ref{ass:selected-deviation}, which are stated in Section~\ref{ssec:tracking}, hold with $\kappa_\pi=0$ and $b_\pi=b_{\mathrm{alt}}=r_u$.
\begin{remark}[Effective sample size]
With normalized eligible weights, $\mathrm{ESS}_k:=1/\sum_b\omega_b^2$ and $\widehat\Sigma_{u,k}:=\sum_b\omega_b(u_k^{(b)}-\widehat u_k^{(N_f)})(u_k^{(b)}-\widehat u_k^{(N_f)})^\top$. The quantity $\sqrt{\operatorname{tr}(\widehat\Sigma_{u,k})/\mathrm{ESS}_k}$ is a useful online diagnostic of the variability of the weighted input.
\end{remark}
\subsection{Finite-Sample Stochastic Closed-Loop Stability}
\label{ssec:tracking}
Let $u_k^{\mathrm{fb}}:=\ub_k+K_k(\xb_k\ominus x_k)$ and $\Delta u_k^{\circ,(N_f)}:=u_k^{\circ,(N_f)}-u_k^{\mathrm{fb}}$.
\begin{assumption}[Conditional population-control deviation]
\label{ass:population-deviation}
There exist $\kappa_\pi,b_\pi\geq0$ such that $\|\Delta u_k^{\circ,(N_f)}\|_2\leq\kappa_\pi\|e_k\|_2+b_\pi$ uniformly over the mission steps and reached states.
\end{assumption}
The constants $(\kappa_\pi,b_\pi)$ are bounds that are used in the analysis. A small $\kappa_\pi$ is expected when the tracking cost dominates the mission cost, whereas deliberate avoidance maneuvers are represented by the offset $b_\pi$.
\begin{assumption}[Alternative executed-action deviation]
\label{ass:selected-deviation}
Whenever the weighted input does not satisfy~\eqref{eq:executed-dcbf-test}, any sampled or backup action executed by Algorithm~\ref{alg:csmc} satisfies $\|u_k^\star-u_k^{\mathrm{fb}}\|_2\leq b_{\mathrm{alt}}$ for some $b_{\mathrm{alt}}\geq0$.
\end{assumption}
\begin{assumption}[Uniform conditional prediction-error moments]
\label{ass:executed-prediction-moments}
For every executable $\mathcal H_k$-measurable $u\in\mathcal U$, deterministic bounds $\mu_{d,k},\sigma_{d,k}\geq0$ satisfy
\begin{equation}
    \|\bar d_{x,k}(u)\|_{M_{k+1}}^2\leq\mu_{d,k}^2,
    \ \
    \operatorname{tr}(M_{k+1}\Sigma_{d,k}(u))\leq\sigma_{d,k}^2
    \ \text{a.s.}
\end{equation}
\end{assumption}
Let $\rho_{\mathrm{cl}}:=\sqrt\beta+L_u^M\kappa_\pi/\sqrt{\underline m}$. The condition $\rho_{\mathrm{cl}}<1$ requires the state-dependent portion of the conditional population action to remain within the nominal contraction margin.
In the following theorem, we propagate the sampling error of Lemma~\ref{lem:csmc-concentration} and the prediction error through the contracting dynamics, which gives the stochastic stability bound for the implemented controller.
\begin{theorem}[Finite-sample safety and stability]
\label{thm:tracking}
Suppose Proposition~\ref{prop:tubewise} and Assumptions~\ref{ass:input-sensitivity},~\ref{ass:stochastic-prediction-error},~\ref{ass:weighted-moments},~\ref{ass:population-deviation},~\ref{ass:selected-deviation}, and~\ref{ass:executed-prediction-moments} hold. Assume the executed state remains in the certified domain and coordinate chart, $u_k^{\mathrm{fb}}\in\mathcal U$ on the reached states, and $\rho_{\mathrm{cl}}<1$.

Fix $\eta_s\in(0,1)$ and $N_f\geq N_0(\eta_s)$. Choose $\eta_u,\eta_b,\eta_d>0$ such that
\begin{equation}
    \beta_{\mathrm{fs}}:=(1+\eta_d)(1+\eta_u)\rho_{\mathrm{cl}}^2<1.
    \label{eq:finite_sample_stochastic_rate}
\end{equation}
Let $b_\star:=\max\{b_\pi,b_{\mathrm{alt}}\}$ and define
\begin{subequations}
\begin{align}
    c_{\mathrm{alg},2}
    &:={}
    (1+\eta_d)(1+\eta_u^{-1})(L_u^M)^2
    \Bigl((1+\eta_b)b_\star^2
    \nonumber\\
    &\hspace{5em}+(1+\eta_b^{-1})
    \varepsilon_{\mathrm{samp}}^2(N_f,\eta_s)\Bigr),
    \label{eq:algorithmic_second_moment_floor}
    \\
    c_{d,k}^{\mathrm{exec}}
    &:={}
    (1+\eta_d^{-1})\mu_{d,k}^2+\sigma_{d,k}^2.
    \label{eq:executed_stochastic_floor}
\end{align}
\end{subequations}
Then
\begin{equation}
    \E[\mathcal V_{k+1}\mid\mathcal F_k]
    \leq\beta_{\mathrm{fs}}\mathcal V_k+c_{\mathrm{alg},2}+c_{d,k}^{\mathrm{exec}}.
    \label{eq:finite_sample_conditional_metric_recursion}
\end{equation}
Consequently,
\begin{equation}
    \E[\mathcal V_k]
    \leq\beta_{\mathrm{fs}}^k\E[\mathcal V_0]
    +\sum_{j=0}^{k-1}\beta_{\mathrm{fs}}^{k-1-j}
      (c_{\mathrm{alg},2}+c_{d,j}^{\mathrm{exec}}).
    \label{eq:finite_sample_metric_bound}
\end{equation}
If $c_{d,k}^{\mathrm{exec}}\leq\overline c_d^{\mathrm{exec}}$ uniformly, then
\begin{equation}
    \E[\mathcal V_k]
    \leq\beta_{\mathrm{fs}}^k\E[\mathcal V_0]
    +(c_{\mathrm{alg},2}+\overline c_d^{\mathrm{exec}})
      \frac{1-\beta_{\mathrm{fs}}^k}{1-\beta_{\mathrm{fs}}}.
    \label{eq:uniform_finite_sample_metric_bound}
\end{equation}
The Euclidean mean-square error satisfies
\begin{equation}
\begin{aligned}
    \E[\|e_k\|_2^2]
    \leq{}&\frac{\overline m}{\underline m}\,
    \beta_{\mathrm{fs}}^k\E[\|e_0\|_2^2]
    \\
    &+\frac{1}{\underline m}
    \sum_{j=0}^{k-1}\beta_{\mathrm{fs}}^{k-1-j}
    (c_{\mathrm{alg},2}+c_{d,j}^{\mathrm{exec}}).
\end{aligned}
\label{eq:finite_sample_state_mse_bound}
\end{equation}
If, additionally, $h(x_0,t_0)\geq0$, Assumption~\ref{ass:backup} holds, Algorithm~\ref{alg:csmc} applies~\eqref{eq:executed-dcbf-test} at every step, and~\eqref{eq:method_risk_allocation} holds, then
\begin{equation}
    \Pr\!\left(h(x_k,t_k)\geq0,\ k=0,\ldots,N\right)
    \geq1-\delta_{\mathrm{mis}}.
    \label{eq:main_executed_chance_safety}
\end{equation}
\end{theorem}
\begin{IEEEproof}
Let $\xi_k:=\|\widehat u_k^{(N_f)}-u_k^{\circ,(N_f)}\|_2$. If the weighted input is executed, Assumption~\ref{ass:population-deviation} gives $\|u_k^\star-u_k^{\mathrm{fb}}\|_2\leq\kappa_\pi\|e_k\|_2+b_\pi+\xi_k$; the alternative branch is bounded by Assumption~\ref{ass:selected-deviation}. Therefore, both branches satisfy
\begin{equation}
    \|u_k^\star-u_k^{\mathrm{fb}}\|_2
    \leq\kappa_\pi\|e_k\|_2+b_\star+\xi_k.
    \label{eq:unified_action_deviation}
\end{equation}
Define $\widetilde e_{k+1}:=f(x_k,u_k^\star)\ominus\xb_{k+1}$. Proposition~\ref{prop:tubewise}, Assumption~\ref{ass:input-sensitivity}, and~\eqref{eq:unified_action_deviation} imply
\begin{equation}
    \|\widetilde e_{k+1}\|_{M_{k+1}}
    \leq\rho_{\mathrm{cl}}\sqrt{\mathcal V_k}
    +L_u^M(b_\star+\xi_k).
    \label{eq:nominal_executed_error_bound}
\end{equation}
Since $e_{k+1}=\widetilde e_{k+1}+d_{x,k}(u_k^\star)$, conditioning on $\mathcal H_k$ gives
\begin{equation}
\begin{aligned}
    \E[\mathcal V_{k+1}\mid\mathcal H_k]
    \leq{}&(1+\eta_d)
    \|\widetilde e_{k+1}\|_{M_{k+1}}^2
    \\
    &+(1+\eta_d^{-1})\mu_{d,k}^2+\sigma_{d,k}^2.
\end{aligned}
\label{eq:post_sampling_stochastic_recursion}
\end{equation}
Applying Young's inequality to~\eqref{eq:nominal_executed_error_bound} and $(b_\star+\xi_k)^2$, and then using~\eqref{eq:conditional_sampling_mse}, yields~\eqref{eq:finite_sample_conditional_metric_recursion}. The tower property, iteration, and metric norm equivalence give the stated mean-square bounds. The probabilistic safety result follows from Proposition~\ref{prop:executed-safety}.
\end{IEEEproof}
\begin{remark}[Interpretation and scope]
Theorem~\ref{thm:tracking} separates three effects. The population correction $\kappa_\pi$ changes the contraction rate from $\sqrt\beta$ to $\rho_{\mathrm{cl}}$. The offsets $b_\pi$ and $b_{\mathrm{alt}}$ and the finite-sample error enter the floor $c_{\mathrm{alg},2}$, and the prediction-error moments enter the floor $c_{d,k}^{\mathrm{exec}}$. Increasing $N_f$ reduces the finite-sample part of the floor. The bounds hold until the first exit from the certified domain, and the Markov-based planning margin in Remark~\ref{rem:tighten} can be used to keep the exit probability small. For bounded disturbances, Proposition~\ref{prop:bounded-validation} gives an invariant domain.
\end{remark}
\begin{proposition}[Invariant domain under bounded disturbances]
\label{prop:bounded-validation}
In a Euclidean chart, let $\mathcal D_k:=\{x\mid\|x-\xb_k\|_{M_k}\leq\varrho\}$ be contained in the certified domain $\mathcal X$, with $u_k^{\mathrm{fb}}(x)\in\mathcal U$ for $x\in\mathcal D_k$. Suppose that every executed input, including every fallback, satisfies $\|u_k^\star-u_k^{\mathrm{fb}}(x_k)\|_2\leq r_u$, and that the total prediction error satisfies $\|d_{x,k}(u)\|_{M_{k+1}}\leq d_{\max}$ for every $u\in\mathcal U$ and every realization. If
\begin{equation}
    \sqrt\beta\,\varrho+L_u^Mr_u+d_{\max}\leq\varrho,
    \label{eq:invariant-support-test}
\end{equation}
then $x_0\in\mathcal D_0$ implies $x_k\in\mathcal D_k$ for $k=0,\ldots,N$. Furthermore, the feedback input $u_k^{\mathrm{fb}}(x_k)$ satisfies~\eqref{eq:executed-dcbf-test} with the barrier $h(x,t_k)=\varrho-\|x-\xb_k\|_{M_k}$, the rate $\alpha=1$, and the region $\mathcal E_k=\{d\mid\|d\|_{M_{k+1}}\leq d_{\max}\}$, which has conditional coverage one.
\end{proposition}
\begin{IEEEproof}
Using~\eqref{eq:reference_contraction_bound}, Assumption~\ref{ass:input-sensitivity}, and the triangle inequality, we obtain $\|e_{k+1}\|_{M_{k+1}}\leq\sqrt\beta\|e_k\|_{M_k}+L_u^Mr_u+d_{\max}$, and the first claim follows from~\eqref{eq:invariant-support-test} by induction. For the feedback input, the barrier $h$ is $1$-Lipschitz in the metric norm, so that $\Delta_{h,k}^{\delta_k}\leq d_{\max}$ and $h(F_k(x_k),t_{k+1})-d_{\max}\geq\varrho-\sqrt\beta\varrho-d_{\max}\geq0$, which is~\eqref{eq:executed-dcbf-test} with $\alpha=1$.
\end{IEEEproof}
Proposition~\ref{prop:bounded-validation} gives a setting in which all the hypotheses of Theorem~\ref{thm:tracking} are verified before the simulation: the state remains in the certified domain, the residual clipping of Corollary~\ref{cor:shifted-concentration} gives $\kappa_\pi=0$ and $b_\pi=b_{\mathrm{alt}}=r_u$, the support region has conditional coverage one, and the feedback input is a feasible backup. We use this setting in Section~\ref{ssec:direct-validation} to evaluate the bounds numerically.
For conditionally centered prediction error, one may set $\beta_{\mathrm{fs}}^0:=(1+\eta_u)\rho_{\mathrm{cl}}^2$, $c_{d,k}^{\mathrm{exec},0}:=\sigma_{d,k}^2$, and omit the factor $1+\eta_d$ from~\eqref{eq:algorithmic_second_moment_floor}. If the assumptions hold uniformly for all $k\geq0$, then
\begin{equation}
    \limsup_{k\to\infty}\E[\|e_k\|_2^2]
    \leq\frac{c_{\mathrm{alg},2}+\overline c_d^{\mathrm{exec}}}
    {\underline m(1-\beta_{\mathrm{fs}})}.
    \label{eq:executed_state_mean_square_limsup}
\end{equation}
For a target root-mean-square (RMS) input-approximation accuracy $\varepsilon_{\mathrm{tr}}>0$, choose $\eta_s$ such that $\eta_sd_{\mathcal U}^2<\varepsilon_{\mathrm{tr}}^2$. It is sufficient to take
\begin{equation}
    N_f\geq N_{\mathrm{tr}}:=\max\!\left\{
        N_0(\eta_s),
        \left\lceil\frac{C_u^2\log(4m/\eta_s)}
        {\varepsilon_{\mathrm{tr}}^2-\eta_sd_{\mathcal U}^2}\right\rceil
    \right\}.
    \label{eq:tracking-sample-budget}
\end{equation}
Equivalently, $N_s\geq N_{\mathrm{tr}}+3$, which ensures $\varepsilon_{\mathrm{samp}}^2(N_f,\eta_s)\leq\varepsilon_{\mathrm{tr}}^2$. The constant $C_u$ scales as $q_{\min}^{-1}$, so this worst-case requirement is conservative when the importance weights are highly concentrated, and we evaluate the sample budgets empirically in Section~\ref{sec:sims}.
\subsection{Role of the Sample Budget in Avoidance}
\label{ssec:avoidance-sample-budget}
Once Assumption~\ref{ass:backup} and the conditional quantile model are available, Proposition~\ref{prop:executed-safety} holds for any number of free rollouts. The sample budget controls the approximation of the population input and the speed with which the warm-start mean finds a progress-preserving maneuver. Therefore, Theorem~\ref{thm:tracking} is the formal finite-sample result, while the maneuver discovery is evaluated empirically in Section~\ref{sec:sims} as a function of $N_s$.
\begin{remark}[Optional stochastic planning margin]
\label{rem:tighten}
Let $B_k$ denote the right-hand side of~\eqref{eq:finite_sample_state_mse_bound}. Markov's inequality gives $\Pr(\|e_k\|_2\leq r_{x,k}^{\mathrm{plan}})\geq1-\delta_{\mathrm{plan},k}$ for $r_{x,k}^{\mathrm{plan}}:=\sqrt{B_k/\delta_{\mathrm{plan},k}}$. If an available planning constraint $h_{\mathrm{plan}}(\cdot,t_k)$ is $L_{\mathrm{plan}}$-Lipschitz, imposing
\begin{equation}
    h_{\mathrm{plan}}(x_k^{\mathrm{plan}},t_k)
    \geq L_{\mathrm{plan}}r_{x,k}^{\mathrm{plan}}
\end{equation}
provides a per-step confidence of at least $1-\delta_{\mathrm{plan},k}$ for the corresponding Lipschitz planning constraint. This optional horizon-wide margin is distinct from the one-step executed-input tightening.
\end{remark}
\section{Simulation Studies}
\label{sec:sims}
In this section, we validate CS-MPC on \rev{three nonlinear systems: 1) a skid-steer rover tracking a path among moving nonconvex obstacles under terrain-slip uncertainty, which tests the sample efficiency of the feedback-closed rollouts, 2) a quadrotor tracking a figure-eight among moving obstacles, which tests the execution layer on an unstable system, and 3) a spacecraft landing on an asteroid} with minimum-impulse thrusters under uncertain gravity and spin, which tests a nonsmooth fuel cost under persistent parametric uncertainty. For the asteroid, we verify the contraction condition analytically over a domain that contains the executed trajectories. For the rover and the quadrotor, the contraction condition is validated by sampling near the reference, and the trajectories leave this domain during the avoidance maneuvers. Finally, we evaluate the bounds of Section~\ref{sec:main-results} on a stationkeeping experiment, in which the spacecraft holds a fixed position near the asteroid under bounded disturbances.

\subsection{Simulation Setup}
We compare CS-MPC with two versions of the standard MPPI controller of Definition~\ref{def:mppi}, which use the same horizon, covariance, and temperature, with open-loop rollouts around the warm-started mean sequence, which is kept in $\mathcal U$, and with the weighted input applied directly. In the first version (MPPI$^{\mathrm o}$), the safety constraints enter the rollout cost as obstacle penalties. In the second version (MPPI$^{\mathrm b}$), the rollout cost also includes the barrier penalty $J_{\mathrm{DCBF},k}$ in~\eqref{eq:cost}, which gives the rollout cost of CS-MPC, and the figures show this version. The risk is allocated uniformly as $\delta_k=\delta_{\mathrm{mis}}/N$ with $\delta_{\mathrm{mis}}=0.05$, and the quantile radius is computed from $N_{\mathrm{cal}}=2\times10^5$ calibration draws as described in Remark~\ref{rem:quantile-construction}. The execution layer uses the lookahead barrier of Remark~\ref{rem:lookahead}, and we report the physical clearance separately. In the implementation, the execution layer accepts an input only if it satisfies~\eqref{eq:executed-dcbf-test} and the nominal rollout, which applies this input followed by the backup policy for $H_b=10$ steps, keeps $h\geq L_hr_k^{\delta_k}$. If the weighted input is not accepted, then the execution layer applies the accepted candidate that is closest to the weighted input, among the sampled inputs and the convex combinations of the weighted input and the backup input, so that the tracking input is modified minimally. If no candidate is accepted, then the backup input is applied when it passes the same check; otherwise, the step is flagged and the candidate with the largest predicted barrier value is applied. Note that this recoverability check verifies the backup feasibility of Assumption~\ref{ass:backup} online for the nominal model. Unless stated otherwise, we run each configuration on five seeds and report the median. We implement the controllers in Python using NumPy, and we ran the simulations on a single core of an Apple silicon M4 processor.

\subsection{Example 1: Rover Among Moving Obstacles Under Terrain Slip}
\label{ssec:sim-rover}
We consider a skid-steer rover with the state $x=[p_x,p_y,\theta,\omega_{FL},\omega_{FR},\omega_{RL},\omega_{RR}]^\top\in\R^7$, where $(p_x,p_y)$ is the position, $\theta$ is the heading, and $\omega_j$ are the wheel speeds, and with the four wheel-speed commands $u\in[-6.2,6.2]^4$\,rad/s as the input. The dynamics of the rover are of the form~\eqref{eq:dyn} with $d_{a,k}\equiv0$ and are given as follows:
\begin{equation}
\begin{aligned}
    x_{k+1}&=f(x_k,u_k)+\Delta f(x_k,u_k;\vartheta)+w_k,\\
    f(x,u)&=x+\Delta t
    \begin{bmatrix}
        v_b\cos\theta\\ v_b\sin\theta\\ \Omega_b\\ (u-\omega)/\tau_w
    \end{bmatrix},\\
    \Delta f(x,u;\vartheta)&=\Delta t
    \begin{bmatrix}
        (s_x-1)v_b\cos\theta-s_\ell s_xv_b\sin\theta\\
        (s_x-1)v_b\sin\theta+s_\ell s_xv_b\cos\theta\\
        (s_\theta-1)\Omega_b\\ 0
    \end{bmatrix},
\end{aligned}
\label{eq:rover_dynamics}
\end{equation}
where $\omega=[\omega_{FL},\omega_{FR},\omega_{RL},\omega_{RR}]^\top$, $v_b=\tfrac{r_w}{2}(\omega_R+\omega_L)$ is the nominal forward speed, $\Omega_b=\tfrac{r_w}{b}(\omega_R-\omega_L)$ is the nominal yaw rate, and $\omega_L$ and $\omega_R$ are the mean left and right wheel speeds. We use the wheel radius $r_w=0.065$\,m, the track width $b=0.36$\,m, the wheel time constant $\tau_w=0.15$\,s, and $\Delta t=0.1$\,s. The terrain slip is modeled by the mission parameter $\vartheta=(s_x,s_\theta,s_\ell)$, which is drawn once per run from uniform distributions with the longitudinal traction $s_x\in[0.86,1.0]$, the yaw traction $s_\theta\in[0.90,1.10]$, and the lateral slip fraction $s_\ell\in[-0.04,0.04]$, so that $\Delta f=0$ for the nominal value $\vartheta=(1,1,0)$. The process noise is $w_k\sim\mathcal N(0,W)$ with per-step standard deviations of $0.004$\,m, $0.004$\,rad, and $0.08$\,rad/s in the position, heading, and wheel speeds.

\begin{figure}[!t]
\centering
\includegraphics[width=0.6\columnwidth]{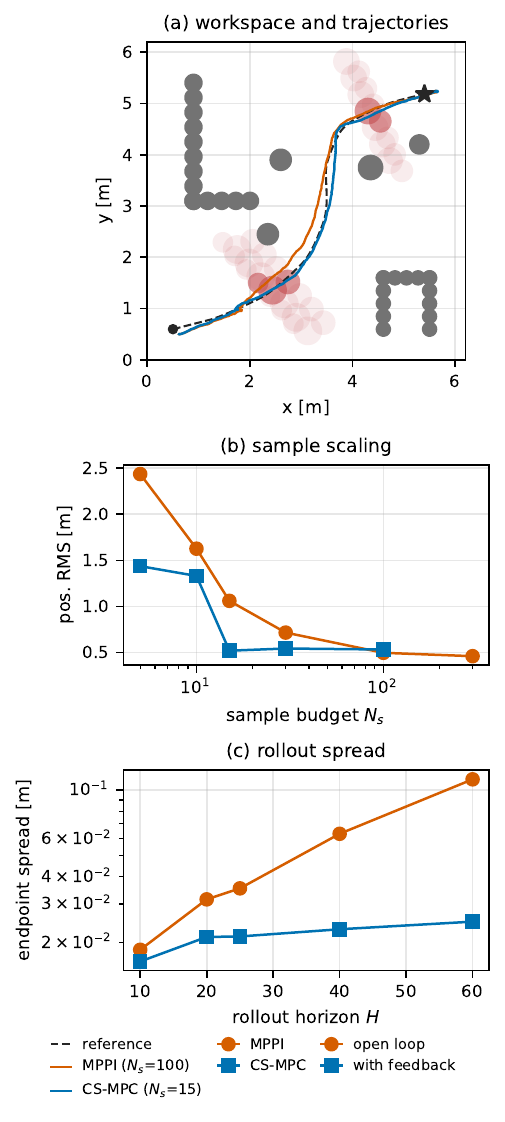}
\caption{We compare CS-MPC and MPPI for the rover. (a) The workspace with the static obstacles, the sweep of the two moving obstacles (red), and the executed trajectories. (b) The position RMS for different sample budgets. (c) The endpoint spread of the rollouts for different horizons.}
\label{fig:sim-rover}
\end{figure}

The workspace shown in Fig.~\ref{fig:sim-rover}(a) has static obstacles, which define $h_s$, and two nonconvex obstacles, which define $h_d$ and oscillate across the reference path with a peak speed of $0.25$\,m/s that is close to the $0.26$\,m/s cruise speed of the rover. The reference is a $34$\,s rollout of the nominal model ($N=340$), for which the sampled Jacobian checks give $\beta=0.967$ near the reference, and the backup policy moves the rover along the clearance gradient.

\rev{In Table~\ref{tab:rover} and Fig.~\ref{fig:sim-rover}(b), we compare CS-MPC with the two MPPI versions for different sample budgets. We observe that CS-MPC with $N_s=15$ attains a position RMS of $0.52$\,m, which MPPI attains with $N_s=100$ to $300$, and that CS-MPC has no collisions at any tested budget, whereas both MPPI versions collide for $N_s\leq10$. Over $50$ missions, MPPI$^{\mathrm o}$ has $9$ barrier violations with $N_s=30$ and $4$ barrier violations with $N_s=100$, when compared with no violations for CS-MPC with $N_s=15$. Without the execution layer, the feedback-closed rollouts attain a position RMS of $0.33$--$0.37$\,m but collide on $4$ of $5$ seeds at $N_s=15$, which shows that the execution layer provides the safety at the cost of the avoidance maneuver. As shown in Fig.~\ref{fig:sim-rover}(c), the endpoint spread of the feedback-closed rollouts grows from $0.016$\,m to $0.025$\,m when the horizon increases from $H=10$ to $60$, while the open-loop spread grows from $0.018$\,m to $0.112$\,m, which is consistent with Proposition~\ref{prop:concentration}.}
\begin{table}[!t]
\centering
\caption{Rover sample scaling (median over $5$ seeds). MPPI$^{\mathrm o}$: obstacle penalties; MPPI$^{\mathrm b}$: obstacle and barrier penalties. RMS: position RMS [m]; Clr.: minimum physical clearance over seeds [m]; Coll.: colliding seeds; Step: mean step time [ms].}
\label{tab:rover}
\footnotesize\setlength{\tabcolsep}{5pt}

\begin{tabular}{lrrrrr}
\toprule
Method & $N_s$ & RMS & Clr. & Coll. & Step \\
\midrule
\textbf{CS-MPC} & 5 & 1.437 & 0.100 & 0/5 & 27.0 \\
\textbf{CS-MPC} & 10 & 1.331 & 0.101 & 0/5 & 27.5 \\
\textbf{CS-MPC} & 15 & 0.521 & 0.090 & 0/5 & 25.1 \\
\textbf{CS-MPC} & 30 & 0.545 & 0.109 & 0/5 & 25.4 \\
\textbf{CS-MPC} & 100 & 0.534 & 0.112 & 0/5 & 27.9 \\
\midrule
MPPI$^{\mathrm b}$ & 5 & 2.436 & $-0.450$ & 5/5 & 7.8 \\
MPPI$^{\mathrm b}$ & 10 & 1.627 & $-0.235$ & 2/5 & 8.3 \\
MPPI$^{\mathrm b}$ & 15 & 1.061 & 0.086 & 0/5 & 8.1 \\
MPPI$^{\mathrm b}$ & 30 & 0.717 & 0.092 & 0/5 & 8.7 \\
MPPI$^{\mathrm b}$ & 100 & 0.498 & 0.093 & 0/5 & 10.2 \\
MPPI$^{\mathrm b}$ & 300 & 0.462 & 0.094 & 0/5 & 15.2 \\
\midrule
MPPI$^{\mathrm o}$ & 5 & 2.107 & $-0.359$ & 5/5 & 7.8 \\
MPPI$^{\mathrm o}$ & 10 & 1.714 & $-0.152$ & 1/5 & 8.1 \\
MPPI$^{\mathrm o}$ & 15 & 1.046 & 0.092 & 0/5 & 8.5 \\
MPPI$^{\mathrm o}$ & 30 & 0.952 & $-0.168$ & 1/5 & 8.4 \\
MPPI$^{\mathrm o}$ & 100 & 0.503 & 0.090 & 0/5 & 10.3 \\
MPPI$^{\mathrm o}$ & 300 & 0.455 & 0.095 & 0/5 & 15.2 \\
\bottomrule
\end{tabular}

\end{table}

\subsection{Example 2: Quadrotor Among Moving Obstacles}
\label{ssec:sim-quad}
We consider a quadrotor with the state $x=[p,v,\eta,\omega]\in\R^{12}$, where $p$ is the position, $v$ is the velocity, $\eta$ is the attitude in ZYX Euler angles, and $\omega$ is the body rate. The input is $u=[T,\tau]$, with the thrust $T\in[0,30]$\,N and the body torques $\tau\in[-0.6,0.6]^3$\,N\,m. The dynamics of the quadrotor are of the form~\eqref{eq:dyn} with $\Delta f\equiv0$ and $d_{a,k}\equiv0$ and are given as follows:
\begin{equation}
\begin{aligned}
    x_{k+1}&=f(x_k,u_k)+w_k,\\
    f(x,u)&=x+\Delta t
    \begin{bmatrix}
        v\\ -g_0e_3+\tfrac{T}{m_q}C(\eta)e_3\\ S(\eta)\omega\\
        \mathcal I_q^{-1}(-\omega\times\mathcal I_q\omega+\tau)
    \end{bmatrix},
\end{aligned}
\label{eq:quad_dynamics}
\end{equation}
where $m_q=1.0$\,kg is the mass, $\mathcal I_q=\operatorname{diag}(0.01,0.01,0.02)$\,kg\,m$^2$ is the inertia matrix, $g_0=9.81$\,m/s$^2$, $e_3$ is the vertical unit vector, $C(\eta)$ is the rotation matrix from the body frame to the inertial frame, and $S(\eta)$ maps the body rate to the Euler-angle rates. We use $\Delta t=50$\,ms, and the body rates are clipped to $\pm25$\,rad/s after each step. The process noise is $w_k\sim\mathcal N(0,W)$ with per-step standard deviations of $0.005$\,m, $0.03$\,m/s, and $0.06$\,rad/s in the position, velocity, and body rates.

The quadrotor tracks a $12$\,s figure-eight ($N=240$), while two spheres of radius $0.25$\,m oscillate across the reference path. The Riccati design gives a centerline factor $\beta^{\mathrm{ref}}=0.9815$, and the backup policy tilts the thrust away from the nearest obstacle.

\begin{figure}[!t]
\centering
\includegraphics[width=1\columnwidth]{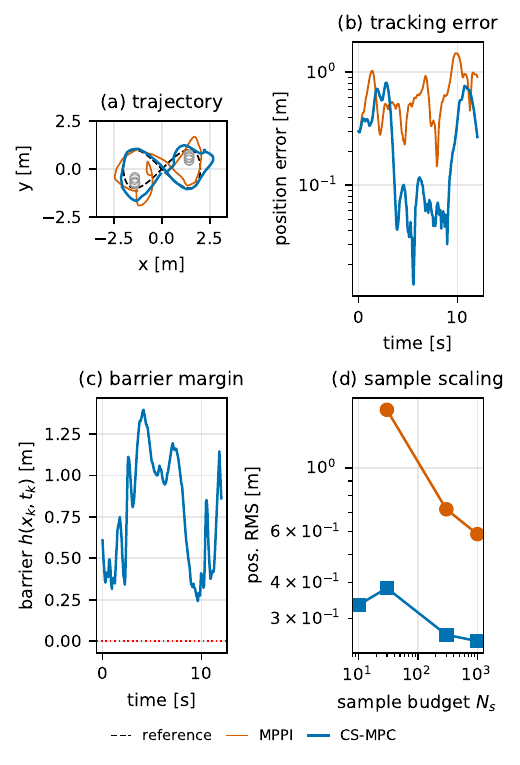}
\caption{We compare CS-MPC ($N_s=30$) and MPPI ($N_s=300$) for the quadrotor. (a) The figure-eight tracking among the moving obstacles. (b) The position error. (c) The executed barrier $h(x_k,t_k)$ of CS-MPC. (d) The position RMS for different sample budgets.}
\label{fig:sim-quad}
\end{figure}

\rev{In Table~\ref{tab:quad} and Fig.~\ref{fig:sim-quad}, we compare CS-MPC with the two MPPI versions. We observe that CS-MPC attains a position RMS of $0.25$--$0.38$\,m for all budgets from $N_s=10$ to $1000$ with no collisions, which is $2$ to $4$ times lower than MPPI with the same number of rollouts. When we remove the execution layer at $N_s=30$, the RMS decreases from $0.38$\,m to $0.17$\,m and the clearance decreases from $0.23$\,m to $0.13$\,m, which is the cost of the avoidance maneuvers for safety.}
\begin{table}[!t]
\centering
\caption{Quadrotor sample scaling (median over $5$ seeds); columns as in Table~\ref{tab:rover}.}
\label{tab:quad}
\footnotesize\setlength{\tabcolsep}{5pt}

\begin{tabular}{lrrrrr}
\toprule
Method & $N_s$ & RMS & Clr. & Coll. & Step \\
\midrule
\textbf{CS-MPC} & 10 & 0.334 & 0.188 & 0/5 & 4.1 \\
\textbf{CS-MPC} & 30 & 0.382 & 0.225 & 0/5 & 4.3 \\
\textbf{CS-MPC} & 300 & 0.263 & 0.146 & 0/5 & 6.9 \\
\textbf{CS-MPC} & 1000 & 0.250 & 0.208 & 0/5 & 13.7 \\
\midrule
MPPI$^{\mathrm b}$ & 30 & 1.594 & 0.379 & 0/5 & 2.7 \\
MPPI$^{\mathrm b}$ & 300 & 0.718 & 0.210 & 0/5 & 5.5 \\
MPPI$^{\mathrm b}$ & 1000 & 0.589 & 0.299 & 0/5 & 12.4 \\
\midrule
MPPI$^{\mathrm o}$ & 30 & 1.134 & 0.116 & 0/5 & 2.7 \\
MPPI$^{\mathrm o}$ & 300 & 0.659 & 0.103 & 0/5 & 5.5 \\
MPPI$^{\mathrm o}$ & 1000 & 0.575 & 0.117 & 0/5 & 12.5 \\
\bottomrule
\end{tabular}

\end{table}

\subsection{Example 3: Asteroid Landing With Minimum-Impulse Thrusters}
\label{ssec:sim-asteroid}
We consider a spacecraft that lands on asteroid (101955) Bennu, for which we use the shape model in~\cite{nolan2013}. The state is $x=[r,v]\in\R^6$, where $r$ and $v$ are the position and velocity in the asteroid-fixed rotating frame, and the input is the commanded acceleration $u\in[-0.002,0.002]^3$\,m/s$^2$. Let $a(x;\vartheta):=-2\omega_a\times v-\omega_a\times(\omega_a\times r)+a_g(r;\vartheta)$ denote the acceleration due to the rotating frame and the gravity, where $\omega_a=[0,0,\omega_a^z]^\top$ is the spin vector of the asteroid. The dynamics of the spacecraft are of the form~\eqref{eq:dyn} and are given as follows:
\begin{equation}
\begin{aligned}
    x_{k+1}&=f(x_k,u_k)+\Delta f(x_k;\vartheta)+d_{a,k}(u_k)+w_k,\\
    f(x,u)&=x+\Delta t
    \begin{bmatrix} v\\ a(x;\bar\vartheta)+u \end{bmatrix},\\
    \Delta f(x;\vartheta)&=\Delta t
    \begin{bmatrix} 0\\ a(x;\vartheta)-a(x;\bar\vartheta) \end{bmatrix},\\
    d_{a,k}(u)&=\Delta t
    \begin{bmatrix} 0\\ Q(u)-u \end{bmatrix},
\end{aligned}
\label{eq:asteroid_dynamics}
\end{equation}
where $\bar\vartheta$ is the nominal value of the mission parameter and $Q$ is the thruster quantizer. The gravitational acceleration is given as follows:
\begin{equation}
\begin{aligned}
    a_g(r;\vartheta)={}&-\frac{\mu r}{\|r\|^3}
    \left[1+C_{20}\frac{R_0^2}{\|r\|^2}
    \left(\frac{3r_z^2}{2\|r\|^2}-\frac12\right)\right]\\
    &-\mu_m\frac{r-r_m}{\|r-r_m\|^3},
\end{aligned}
\label{eq:asteroid_gravity}
\end{equation}
where $\mu$ is the gravitational parameter, $C_{20}$ is the oblateness coefficient, $R_0=246$\,m is the mean radius, $r_z$ is the component of $r$ along the spin axis, and $(\mu_m,r_m)$ are the gravitational parameter and the position of a mascon. The thrusters have a minimum commandable acceleration of $u_{\mathrm{mib}}=10^{-4}$\,m/s$^2$, so that $Q(u)_j=u_j$ if $|u_j|>u_{\mathrm{mib}}$ and $Q(u)_j=0$ otherwise, and we use $\Delta t=10$\,s. The mission parameter is $\vartheta=(\mu,\omega_a^z,C_{20},\mu_m,r_m)$, with the nominal value $\bar\vartheta$ given by $\mu=4.89$\,m$^3$/s$^2$, a $4.296$\,h spin period, $C_{20}=-0.10$, and $\mu_m=0$. The parameter $\vartheta$ is drawn once per run from uniform distributions, with $\mu$ within $\pm25\%$, the spin rate within $\pm15\%$, and $C_{20}$ within $\pm100\%$ of the nominal values, and with $\mu_m$ between $10\%$ and $30\%$ of the nominal $\mu$ at $\|r_m\|=0.7R_0$. The process noise is $w_k\sim\mathcal N(0,W)$ with per-step standard deviations of $0.05$\,m in position and $5\times10^{-4}$\,m/s in velocity.

The reference is a quintic descent from a hold point at $500$\,m altitude over $60$\,min, and a keep-out ball of radius $22$\,m is revealed at $t=35$\,min. The rollout cost includes the $\ell_0$-type firing count $c_u(u)=w_{\mathrm{fire}}\,\#\{j:|u_j|>u_{\mathrm{mib}}\}$, which penalizes the number of thruster firings. For this example, we verify the contraction condition analytically and obtain $\beta^{\mathrm{ver}}=0.845$ over a position box of radius $120$\,m. All the executed trajectories stay inside this box with a maximum position error of $14.7$\,m, so that the certified-domain hypothesis of Theorem~\ref{thm:tracking} holds.

\begin{figure}[!t]
\centering
\includegraphics[width=1\columnwidth]{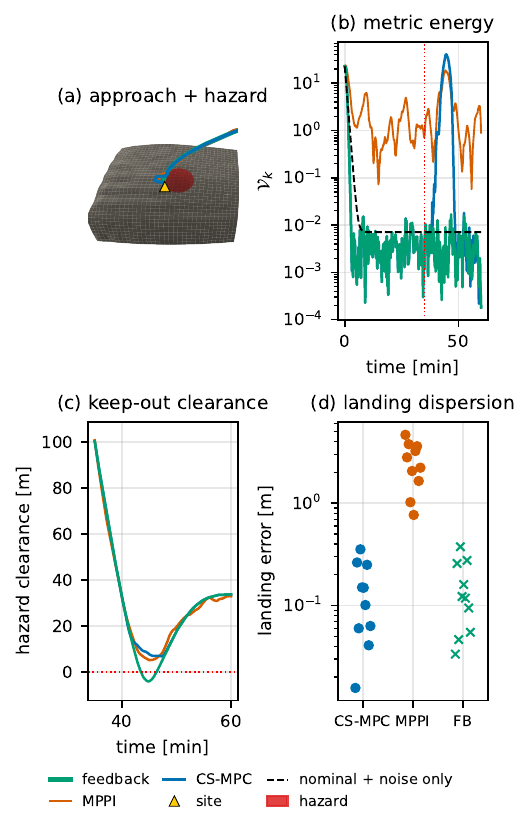}
\caption{We show the results of the asteroid landing. (a) The Bennu shape model~\cite{nolan2013} and the final approach. (b) The metric energy and the nominal envelope of Proposition~\ref{prop:tube}, where the CS-MPC excursion is the avoidance maneuver. (c) The hazard clearance after the reveal. (d) The landing error over the $10$ draws.}
\label{fig:sim-asteroid}
\end{figure}

\rev{In Table~\ref{tab:asteroid} and Fig.~\ref{fig:sim-asteroid}, we compare the controllers over $10$ parameter draws. We observe that CS-MPC clears the hazard on every draw with a minimum clearance of $6.3$\,m and attains a landing error of $0.14$\,m, when compared with $2.6$\,m for MPPI$^{\mathrm b}$, while MPPI$^{\mathrm o}$ and the pure feedback pass through the hazard on all $10$ draws. CS-MPC uses $4.20$\,m/s of $\Delta v$ when compared with $2.14$\,m/s for the pure feedback, where the additional fuel is used for the avoidance maneuver.}
\begin{table}[!t]
\centering
\caption{Asteroid landing over $10$ parameter draws. Land: landing error, mean ($95$th percentile) [m]; TD: touchdown speed [mm/s]; Viol.: hazard violations; Clr.: minimum hazard clearance [m]; Fir.: thruster firings and $\Delta v$ [m/s] (medians).}
\label{tab:asteroid}
\footnotesize\setlength{\tabcolsep}{2pt}

\begin{tabular}{lcccccc}
\toprule
Method & Land & TD & Viol. & Clr. & Fir. & $\Delta v$ \\
\midrule
\textbf{CS-MPC} & 0.14 (0.31) & 1.5 & 0/10 & 6.3 & 740 & 4.20 \\
MPPI$^{\mathrm b}$ & 2.59 (4.28) & 37.2 & 0/10 & 5.1 & 1004 & 11.79 \\
MPPI$^{\mathrm o}$ & 2.44 & -- & 10/10 & $-6.0$ & -- & -- \\
Feedback only & 0.15 (0.33) & 1.3 & 10/10 & $-4.2$ & 689 & 2.14 \\
\bottomrule
\end{tabular}

\end{table}

\subsection{Validation of the Stability and Concentration Bounds}
\label{ssec:direct-validation}
In this subsection, we evaluate the bounds of Theorem~\ref{thm:tracking}, Lemma~\ref{lem:csmc-concentration}, and Proposition~\ref{prop:concentration} on a stationkeeping experiment, for which all the hypotheses are verified and all the constants are computed before the simulation. The spacecraft of Example~3 holds a fixed position at a radius of $746$\,m with $\Delta t=10$\,s, $H=8$, and $N=60$. The parameter ranges of Example~3 are scaled by $0.02$, the process noise is uniform with half-widths of $10^{-3}$\,m and $10^{-5}$\,m/s, the deadband is $10^{-5}$\,m/s$^2$, and the feedforward residuals are clipped to $r_u=5\times10^{-6}$\,m/s$^2$, so that $\kappa_\pi=0$ and $b_\pi=b_{\mathrm{alt}}=r_u$. Using the enclosures in the Appendix, we obtain $\beta=0.793$ on the metric ball of radius $\varrho=0.060$, $L_u^M=76.1$, and $d_{\max}=2.21\times10^{-3}$, so that $\sqrt\beta\varrho+L_u^Mr_u+d_{\max}\approx0.056<\varrho$ and Proposition~\ref{prop:bounded-validation} holds. The rollout cost is at most $0.035$ on this domain, which gives $q_{\min}=0.647$, $N_0=21$, and $\beta_{\mathrm{fs}}=0.858$.

\begin{figure}[!t]
\centering
\includegraphics[width=1\columnwidth]{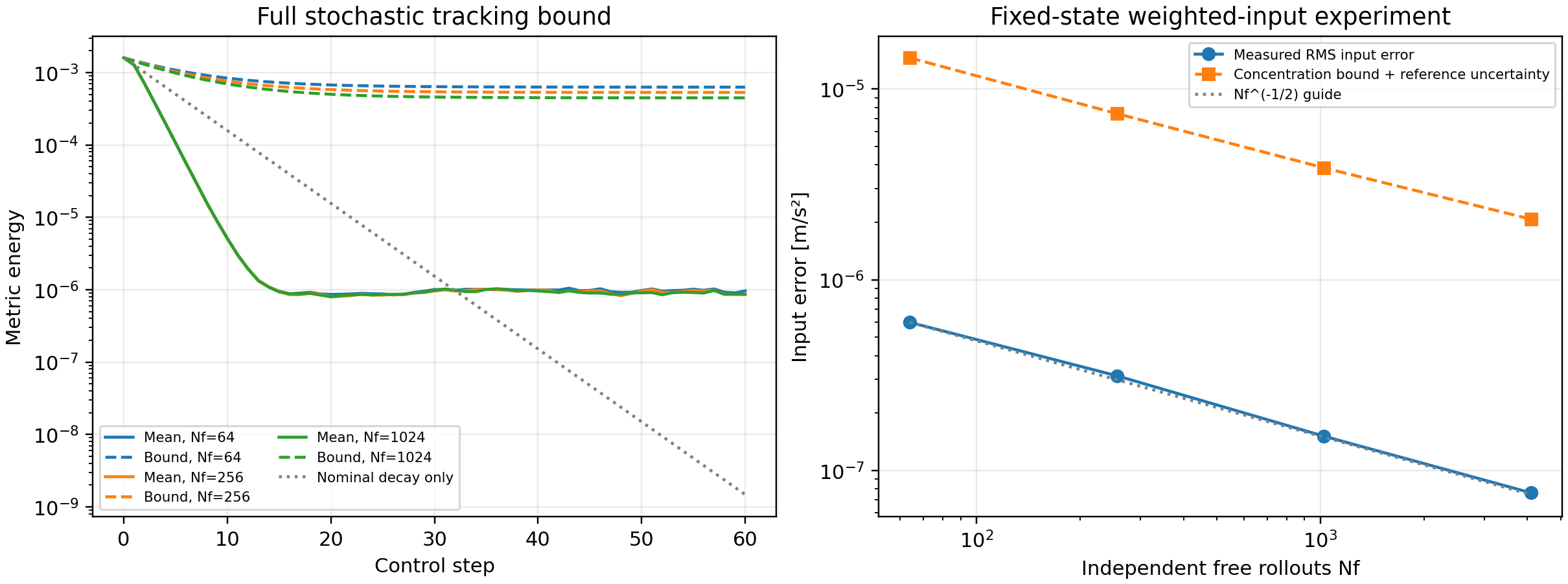}
\caption{We compare the simulated errors with the bounds for the stationkeeping experiment. Left: the mean metric energy over $128$ trials (solid) and the bound of Theorem~\ref{thm:tracking} (dashed). Right: the RMS error of the weighted input (blue), the concentration threshold (orange), and an $N_f^{-1/2}$ guide (dotted).}
\label{fig:theory-validation}
\end{figure}

In Fig.~\ref{fig:theory-validation} (left), we compare the mean metric energy over $128$ Monte Carlo trials with the bound~\eqref{eq:uniform_finite_sample_metric_bound} for $N_f\in\{64,256,1024\}$. We observe that the mean energy remains below the bound for every budget, where the bound is conservative by a factor of approximately $500$ to $650$. For Lemma~\ref{lem:csmc-concentration}, the RMS error of the weighted input over $400$ independent batches decreases with a log--log slope of $-0.498$ for $N_f$ from $64$ to $4096$, which agrees with the $N_f^{-1/2}$ rate, and all the batches are below the threshold of Corollary~\ref{cor:shifted-concentration}. For Proposition~\ref{prop:concentration}, the RMS spread of $4000$ feedback-closed rollouts stays at $1.7\times10^{-3}$ for horizons from $H=10$ to $400$, which is below the bound $1.2\times10^{-2}$ of~\eqref{eq:spread_bound}, while the open-loop spread grows from $1.6\times10^{-2}$ to $8.7$.

\subsection{Violation Rates and Sensitivity}
\label{ssec:sim-checks}
In Table~\ref{tab:safety}, we report the empirical rate at which the barrier $h$ used by the execution layer becomes negative over a larger number of missions. We observe no violations for the three examples, against an allocated risk of $\delta_{\mathrm{mis}}=0.05$, and the execution layer certifies the applied input at more than $99.9\%$ of the steps. Note that these rates are empirical point estimates for the marginal conformal regions of Remark~\ref{rem:quantile-construction}. When we vary the temperature $\gamma\in[0.1,10]$, the horizon $H\in[10,40]$, the exploration scale and the process-noise scale from $0.5$ to $2$ times their nominal values, and the risk budget $\delta_{\mathrm{mis}}\in[0.01,0.5]$ one at a time for the quadrotor at $N_s=30$, the median position RMS over three seeds stays within $0.23$--$0.46$\,m and the minimum clearance is at least $0.13$\,m.
\begin{table}[!t]
\centering
\caption{Empirical safety over $n$ missions (rover $N_s{=}15$, quadrotor $N_s{=}30$, asteroid $N_s{=}20$). $h{<}0$: missions with a barrier violation; Samp./Bkp.: steps that use a sampled or backup input; Inf.: steps flagged by the execution layer.}
\label{tab:safety}
\footnotesize\setlength{\tabcolsep}{3.5pt}

\begin{tabular}{lccccccc}
\toprule
Example & $n$ & $h{<}0$ & Rate & $\delta_{\mathrm{mis}}$ & Samp. & Bkp. & Inf. \\
\midrule
Rover & 50 & 0 & 0.00 & 0.05 & 30.5\% & 0.5\% & 0.0\% \\
Quadrotor & 100 & 0 & 0.00 & 0.05 & 14.9\% & 3.2\% & 0.1\% \\
Asteroid & 50 & 0 & 0.00 & 0.05 & 7.4\% & 0.0\% & 0.0\% \\
\bottomrule
\end{tabular}

\end{table}

\section{Conclusion}
\label{sec:conclusion}
We present a contraction-certified sampling-based MPC method for tracking and safety of nonlinear robotic systems under parametric uncertainty, process disturbances, and safety constraints that are revealed online. The method uses a contraction-certified tracking feedback inside every sampled rollout, which concentrates the rollouts around the reference, and an execution layer that verifies the applied input using a quantile-tightened DCBF condition. We derive an offline verification procedure for nonlinear contraction of the feedback-closed dynamics, and we prove a concentration bound for the weighted control input, a finite-sample stochastic stability bound for the closed-loop system, and a mission-level probabilistic safety guarantee.

We validate CS-MPC in simulation on a rover under terrain-slip uncertainty, a quadrotor flying among moving obstacles, and an asteroid landing with minimum-impulse thrusters. For the rover, CS-MPC with $15$ rollouts attains a tracking error of $0.52$\,m, which standard MPPI control attains with $100$ to $300$ rollouts. For the quadrotor, CS-MPC attains a $2$ to $4$ times lower tracking error than MPPI with the same number of rollouts. For the asteroid landing, CS-MPC attains a landing error of $0.14$\,m when compared with $2.6$\,m for MPPI. CS-MPC has no safety violations in $200$ Monte Carlo missions over the three examples.

\section*{Acknowledgment}
We acknowledge the use of Claude (Anthropic) at multiple stages of this work to edit the text and to improve the Python implementation of the algorithm.

\appendix[Analytic Enclosures for the Stationkeeping Experiment]
\label{app:enclosures}
In this appendix, we derive the constants used in Section~\ref{ssec:direct-validation}. The nominal gravity~\eqref{eq:asteroid_gravity} without the mascon is
\begin{equation}
    a_g(r)=-\mu r\|r\|^{-3}
    -\mu C_{20}R_0^2r\bigl(\tfrac32r_z^2\|r\|^{-7}-\tfrac12\|r\|^{-5}\bigr).
\end{equation}
On $\|r\|\geq r_{\min}>1$, the product rule gives
\begin{equation}
    \|D^2a_g(r)\|\leq\frac{24\mu}{r_{\min}^4}
    +\frac{176\mu|C_{20}|R_0^2}{r_{\min}^6}=:L_G,
\end{equation}
where we use $\|D^2(r\|r\|^{-3})\|\leq24\|r\|^{-4}$, and for $p(r)=rr_z^2$ we use $\|p\|\leq\|r\|^3$, $\|Dp\|\leq3\|r\|^2$, $\|D^2p\|\leq6\|r\|$, $\|D\|r\|^{-7}\|=7\|r\|^{-8}$, and $\|D^2\|r\|^{-7}\|=56\|r\|^{-9}$, which give the coefficient $104$, while the term $r\|r\|^{-5}$ gives $40$, so that $\tfrac32(104)+\tfrac12(40)=176$. The only state-dependent block of the closed-loop Jacobian is $\Delta t\,Da_g(r)$, so that $\Delta_{\Phi,k}\leq\Delta tL_Gr_p$ on a position-error ball of radius $r_p$ with $r_{\min}=\|\bar r\|-r_p$, where $\bar r$ is the reference position, and Corollary~\ref{cor:whitened} gives the contraction factor. The matrix square roots and eigenvalues are computed in floating-point arithmetic.

For the stationary metric, we write $M=L^\top L$, so that the metric ball of radius $\varrho$ has the position radius $r_p=\varrho\|[I\ 0]L^{-1}\|_2$ and the velocity radius $r_v=\varrho\|[0\ I]L^{-1}\|_2$. Let $\mu_0$, $\omega_0$, and $C_{20,0}$ denote the nominal parameters, let $c_\vartheta$ be the scale of the parameter ranges, and let $\Delta\omega=0.15c_\vartheta\omega_0$ and $r_{\max}=\|\bar r\|+r_p$. A uniform bound on the acceleration discrepancy is
\begin{align}
    a_\Delta={}&\frac{0.25c_\vartheta\mu_0}{r_{\min}^2}
    \left(1+|C_{20,0}|(1+c_\vartheta)\frac{R_0^2}{r_{\min}^2}\right)\nonumber\\
    &+\frac{\mu_0|C_{20,0}|c_\vartheta R_0^2}{r_{\min}^4}
    +\frac{0.3c_\vartheta\mu_0}{(r_{\min}-0.7R_0)^2}\nonumber\\
    &+2\Delta\omega r_v
    +(2\omega_0\Delta\omega+\Delta\omega^2)r_{\max},
\end{align}
where the first two terms bound the changes in the gravity parameters, the third term bounds every mascon in the specified support, and the remaining terms bound the changes in the Coriolis and centrifugal accelerations. Let $B=[0;\Delta tI]$ and let the process noise be $w_j\sim\mathrm{Unif}[-\bar w_j,\bar w_j]$ independently. We define
\begin{align}
    d_{\mathrm{bias}}&=\|LB\|_2(a_\Delta+\sqrt3u_{\mathrm{mib}}),\\
    d_w&=\max_{\epsilon\in\{-1,1\}^6}\|L\operatorname{diag}(\bar w)\epsilon\|_2,\\
    d_{\max}&=d_{\mathrm{bias}}+d_w,\qquad\mu_d^2=d_{\mathrm{bias}}^2,\\
    \sigma_d^2&=(\|LB\|_2a_\Delta)^2
    +\operatorname{tr}\!\left(M\operatorname{diag}(\bar w_j^2/3)\right),
\end{align}
where the maximum over the vertices bounds the entire process-noise box by convexity. Given $\mathcal H_k$, the deadband error is deterministic, the contribution of the mission parameter stays in its support ball, and the process noise is independent and zero-mean, which gives the conditional moment bounds uniformly over the invariant domain. Finally, the condition $|\ub_j|+\varrho\|[KL^{-1}]_{j,:}\|_2+r_u\leq u_{\max,j}$ verifies that the feedback and the clipped residual are admissible on the whole ellipsoid.
\bibliographystyle{IEEEtran}
\bibliography{references}
\end{document}